\documentclass[aip,reprint]{revtex4-1}

\usepackage{graphics}
\usepackage{lipsum}
\usepackage[dvipsnames]{xcolor}
\usepackage{hyperref}

\draft 

\begin{document}

\preprint{}

\title{Scalable fabrication of hemispherical solid immersion lenses in silicon carbide through grayscale hard-mask lithography} 



\author{Christiaan Bekker}
\email{c.bekker@hw.ac.uk}

\author{Muhammad Junaid Arshad}
\author {Pasquale Cilibrizzi}
\author{Charalampos Nikolatos}
\affiliation{SUPA, Institute of Photonics and Quantum Sciences, School of Engineering and Physical Sciences, Heriot-Watt University,
Edinburgh EH14 4AS, UK}
\author{Peter Lomax}
\author{Graham S. Wood}
\author{Rebecca Cheung}
\affiliation {Institute for Integrated Micro and Nano Systems, School of Engineering, University of Edinburgh, Scottish Microelectronics Centre, Edinburgh, EH9 3FF, UK}

\author{Wolfgang Knolle}
\affiliation{Leibniz Institute of Surface Engineering (IOM), Permoserstra{\ss}e 15, 04318 Leipzig, Germany}

\author {Neil Ross}
\author {Brian Gerardot}
\author{Cristian Bonato}
\email{c.bonato@hw.ac.uk}
\affiliation{SUPA, Institute of Photonics and Quantum Sciences, School of Engineering and Physical Sciences, Heriot-Watt University,
Edinburgh EH14 4AS, UK}

\date{\today}

\begin{abstract}
Grayscale lithography allows the creation of micrometer-scale features with spatially-controlled height in a process that is fully compatible with standard lithography. Here, solid immersion lenses are demonstrated in silicon carbide using a novel fabrication protocol combining grayscale lithography and hard-mask techniques to allow nearly hemispherical lenses of 5~$\mu$m radius to be etched into the substrate. The technique is highly scalable and compatible with CMOS technology, and device aspect ratios can be tuned after resist patterning by controlling the chemistry of the subsequent dry etch. These results provide a low-cost, high-throughput and industrially-relevant alternative to focused ion beam milling for the creation of high-aspect-ratio, rounded microstructures.
\end{abstract}


\maketitle 

\section{Introduction}
The reliable creation of large-scale and high-aspect-ratio microlens arrays~\cite{cai_microlenses_2021,yuan_fabrication_2018,nussbaum_design_1997}  can impact several areas of photonics and quantum technology. Microlenses are used, for example, to collimate the outputs of vertical-cavity surface-emitting laser (VCSEL) arrays~\cite{tan_high_2020,wang_high_2010} and quantum emitters~\cite{gschrey_highly_2015,bogucki_ultra-long-working-distance_2020,preus_low-divergence_2022,li_wet-etched_2021}, to increase the sensitivity of image sensors by boosting coupling to the device active area~\cite{schlingloff_microlenses_1998, ke_monolithic_2005,tripathi_doublet_2011}, and to increase the efficiency of interconnects for optical transceiver chips~\cite{mccormick_optical_1992,tang_holographic_1996,ballen_off-axis_2000}.

In quantum technology, micrometer-scale solid immersion lenses (SILs) have played a significant role in efficiently extracting single photons from single solid-state quantum emitters~\cite{marseglia_nanofabricated_2011,hadden_strongly_2010,robledo_high-fidelity_2011}. In solid-state matrices, photon collection is usually limited by total internal reflection, which traps most of the emission in the high-index medium. By removing refraction at large angles, SILs can boost collection efficiency by a factor 10-20, as shown for example for the spin/photon interface associated with single nitrogen-vacancy (NV) centers in diamond~\cite{bernien_control_2014}. Embedding NV centers in micro-fabricated SILs has enabled spectacular breakthroughs such as single-shot projective readout of its electronic spin~\cite{robledo_high-fidelity_2011} , the first loophole-free Bell test~\cite{hensen_loophole-free_2015} and the realization of a multi-node quantum network of remote solid-state quantum devices~\cite{pompili_realization_2021,pompili_experimental_2022}. More recently, this technology has also been extended to similar quantum emitters in other materials with better technological maturity, such as silicon carbide~\cite{castelletto_silicon_2022,sardi_scalable_2020, widmann_coherent_2015}.

Conventionally, individual SILs have been fabricated by focused ion beam (FIB) milling~\cite{marseglia_nanofabricated_2011,hadden_strongly_2010,robledo_high-fidelity_2011, widmann_coherent_2015,fu_semiconductor_2002, nagy_high-fidelity_2019}. FIB gives exquisite shape control but is very time-consuming and expensive, with several hours required to mill a single microlens. One alternative to FIB milling which is typically utilized for larger-scale SIL arrays is resist reflow, a lithography-based technique where photoresist pillars are heated past their glass transition temperature, such that the surface tension rounds the resist surface into lens-shaped structures. This method is fast and scalable, and has been used to create lenses in several materials including polymers~\cite{popovic_technique_1988, haselbeck_microlenses_1993, roy_microlens_2009, feng_fabrication_2012, bae_multifocal_2020, zhou_fabrication_2016} glasses~\cite{nussbaum_design_1997,chen_reflow_2008, he_simple_2003} sapphire~\cite{choi_gan_2004, park_refractive_2001}, diamond~\cite{liu_control_2016, liu_large_2016} and more recently silicon carbide~\cite{sardi_scalable_2020}. However the lens profile is constrained by the surface tension of the resist, with shape control only achievable through changing the degree of reflow (through temperature and time of heating), the wettability of the surface~\cite{park_photoresist_2007, jucius_effect_2017} or the dimensions of the precursor resist pillars~\cite{bae_multifocal_2020}. This imposes serious limitations on the tunability and uniformity of lens structures fabricated through resist reflow~\cite{cai_microlenses_2021}. Further, only very shallow (low sagitta compared to radius) SILs have been demonstrated thus far using this technique, possibly due to the collapse of high-aspect-ratio features when the resist is reflown~\cite{kirchner_zep520a_2016}. A second alternative for fabricating densely-packed lens arrays is the indentation of photoresist layers during spin-coating using self-assembled microsphere monolayers~\cite{xu_fabrication_2021}, with the resultant concave structures taking the shape of the microspheres used.s

Here, we present an alternative to these fabrication processes through utilising a novel process based on grayscale lithography~\cite{kirchner_zep520a_2016, grigaliunas_microlens_2016, erjawetz_bend_2022}, where the ability to vary exposure dose spatially is harnessed through use of a low contrast photoresist to create micrometer-scale structures with precisely-controllable heights. We introduce a grayscale hard-mask technique into this process which decouples the etch chemistry of the resist mask and substrate, consequently enabling increased aspect ratios and process robustness over conventional grayscale lithography. These advances allow us to achieve a large degree of shape control over fabricated structures, while maintaining the scalability and speed inherent in lithography-based approaches to microlens fabrication.

\begin{figure} [t]
    \centering
    \includegraphics{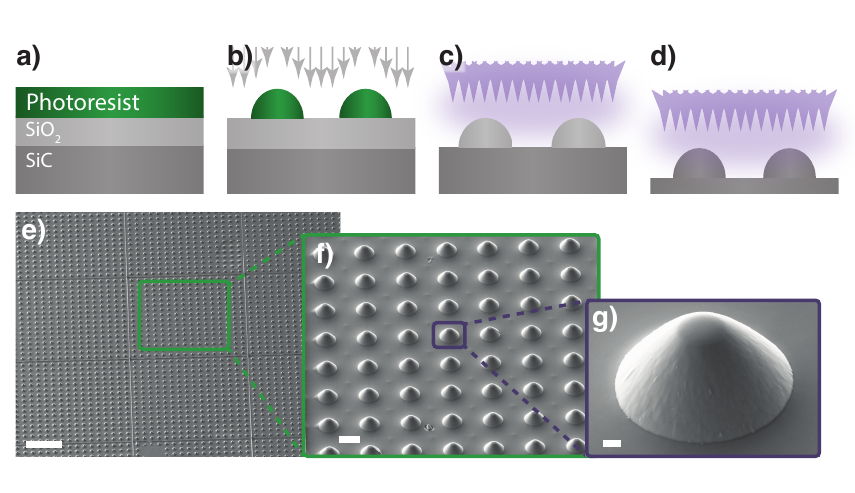}
    \caption{(a-d) Fabrication protocol for hemispherical solid immersion lenses in silicon carbide by grayscale hard-mask lithography, consisting of: (a) spin coating, (b) grayscale exposure and resist development, (c) dry etch of silica layer with photoresist as mask, and (d) dry etch of silicon carbide crystal with silica as a grayscale hard-mask. This process is highly scalable, as demonstrated by SEM micrographs depicting $\sim$1000 (d, scale bar = 100~$\mu$m) and 64 (e, scale bar = 10~$\mu$m) lenses. The resulting lens shape (g, scale bar = 2~$\mu$m) is reproducible across the chip and of the correct dimensions to create 5~$\mu$m-radius hemispherical structures.  }
    \label{Fig1}
\end{figure}

\section{Experimental Details}

\subsection{Device Fabrication Process}

The devices presented here have been fabricated on commercial SiC material (Xiamen PowerWay\textsuperscript{\copyright}, 500~$\mu$m substrate with 15~$\mu$m thick epilayer) diced into 5x5~mm chips. Prior to photolithography, a 5~$\mu$m thick silica (SiO$_2$) layer was deposited on the chip via plasma-enhanced chemical vapor deposition (STS Multiplex CVD). 

Grayscale photoresist (micro resist technology GmbH ma-P 1275G) was spin-coated at 3000rpm to reach a thickness of 8~$\mu$m and soft-baked at a sequence of temperatures up to 100$^\circ$C over 50 minutes. 

The resist was exposed using a Heidelberg Instruments DWL 66+ direct-writing laser lithography system. The exposure beam is stepped over the chip with variable dose, allowing the resist to be partially exposed for spatially-variable height after development. The resist was developed using a TMAH-based solvent (micro resist technology GmbH mr-D 526/S) for 5 minutes, and baked to minimize retained solvent at 60$^\circ$C for 10 minutes.

The chip was subsequently loaded into a reactive ion etching system (RIE; Oxford Instruments PlasmaLab 100), where the SiO$_2$ layer was etched with the grayscale photoresist acting as a mask using a 25/7/3 CHF$_3$/Ar/O$_2$ gas chemistry. In this etching step, the CHF$_3$ etches the SiO$_2$ with high selectivity to the photoresist and the O$_2$ etches the photoresist with high selectivity to the SiO$_2$, allowing for a tuneable total selectivity depending on the desired height of the SiO$_2$ features. The Ar gas acts as a thermal link to improve chip cooling during the etching process, while providing a small amount of physical etching. The etch proceeds until the SiO$_2$ layer has been cleared, ideally coinciding with etching through the photoresist mask layer.

The second etch step focuses on transferring the SiO$_2$ patterns into the SiC substrate. For this RIE step, a SF$_6$/Ar/O$_2$ chemistry is used, where the Ar again acts as a thermal link and the O$_2$ acts both to densify the plasma~\cite{jansen_survey_1996} and to improve the SiC etch rate through chemical etching of carbon~\cite{ahn_study_2004,jiang_inductively_2003,khan_etching_2001,plank_etching_2003}. As SF$_6$ preferentially etches SiC over SiO$_2$ with a selectivity of ~5-10~\cite{oda_optimizing_2015, seok_micro-trench_2020}, this gain allows for a high-selectivity etch step, though this can be tuned through the introduction of CHF$_3$ gas into the etch. 

Utilising a two-step etch process as outlined above brings two benefits to our fabrication protocol: firstly, etching of the photoresist and SiC substrate, which both require oxygen and so cannot be separated in a single etch, is decoupled across two different steps; and secondly, the selectivity of each etch step becomes tuneable through controlling the flow rate of CHF$_3$ gas, which only etches the SiO$_2$. For further discussion see Supplementary Information~\ref{app:Hardmask}. 

The total process time to complete all of the above steps is approximately 12 hours for a pattern including ~20,000 lenses. Of this time 8 hours is spent in etching, limited by the fact that only a conventional RIE system  was available for this process; in principle, an ICP-RIE system can increase the etch rate by up to an order of magnitude~\cite{khan_etching_2001, leerungnawarat_comparison_2001,plank_etching_2003}. Furthermore, the only step in process which has a dependence on the number of lenses is the laser writing step, though writing time can also be reduced by careful consideration of the pattern design. Therefore, the process is immensely scalable and processing at wafer scales with reasonable process times is conceivable.

\subsection{Grayscale Lithography and Pattern Generation}

As mentioned above, grayscale exposure of the photoresist is performed using a Heidelberg Instruments DWL 66+ direct-write laser lithography system. During this step, the laser writer is given an array of grayscale pixels corresponding to an area of the chip, with the brightness of each pixel denoting the dose to be applied at that location of the pattern. Since the ma-p 1275G photoresist used is specifically developed for grayscale lithography, this dose variation results in a resist height variation after exposure. However, the relationship between dose and final resist height is not necessarily linear, and depends on factors such as the intrinsic resist exposure curve, development parameters and baking process prior to exposure.

To calibrate the developed resist height to targeted values, slowly-varying linear slope patterns were exposed in the resist, initially with a linear exposure dose. The slope profiles were then measured using a stylus profilometer (KLA Tencor$^{\text{TM}}$ P-7), both after resist development and after the fabrication process was completed. By comparing the targeted height of the profile for each dose value to the actual height achieved, the dose curve can be calibrated. As this procedure corrects for the universal sensitivity curve of the resist itself, the same calibration was applied to all 3D shapes fabricated, and additional shape modifications arising from the design itself, for example through the local spread of the exposure beam, are addressed through proximity correction (see Section \ref{sec:proximity}).

\begin{figure} [ht]
    \centering
    \includegraphics{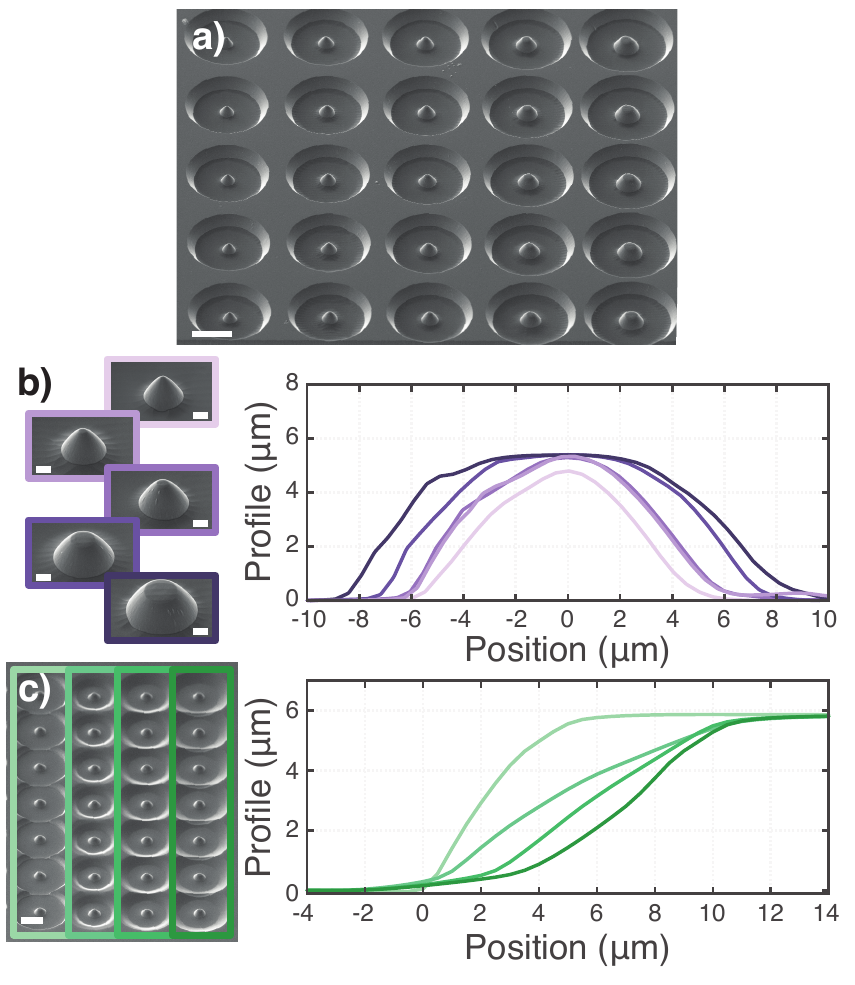}
    \caption{Shape control of features through grayscale lithography. (a) SEM of pattern with varying lens shapes, with profiles of individual lens shapes (b) displaying a progression from pointed conical lenses to wide, flat-topped shapes. The surrounding trenches can also be modified (c) to optimize sidewall angle and shape.}
    \label{Fig2}
\end{figure}

Figure \ref{Fig2} demonstrates the shape control possible through grayscale hardmask lithography. As the shape of grayscale features, in this work rounded solid immersion lenses and trenches, can be arbitrarily modified within fabrication limitations, it is straightforward to create arrays of differently-shaped devices (Fig. \ref{Fig2}(a)), and sweep design parameters such as lens width (Fig. \ref{Fig2}(b)) and trench curvature (Fig. \ref{Fig2}(c)). Therefore the desired shape could be achieved through iterative modification of the design profile, or alternatively through choosing the most desirable shape from a parameter sweep. 

\begin{figure*} [t]
    \centering
    \includegraphics{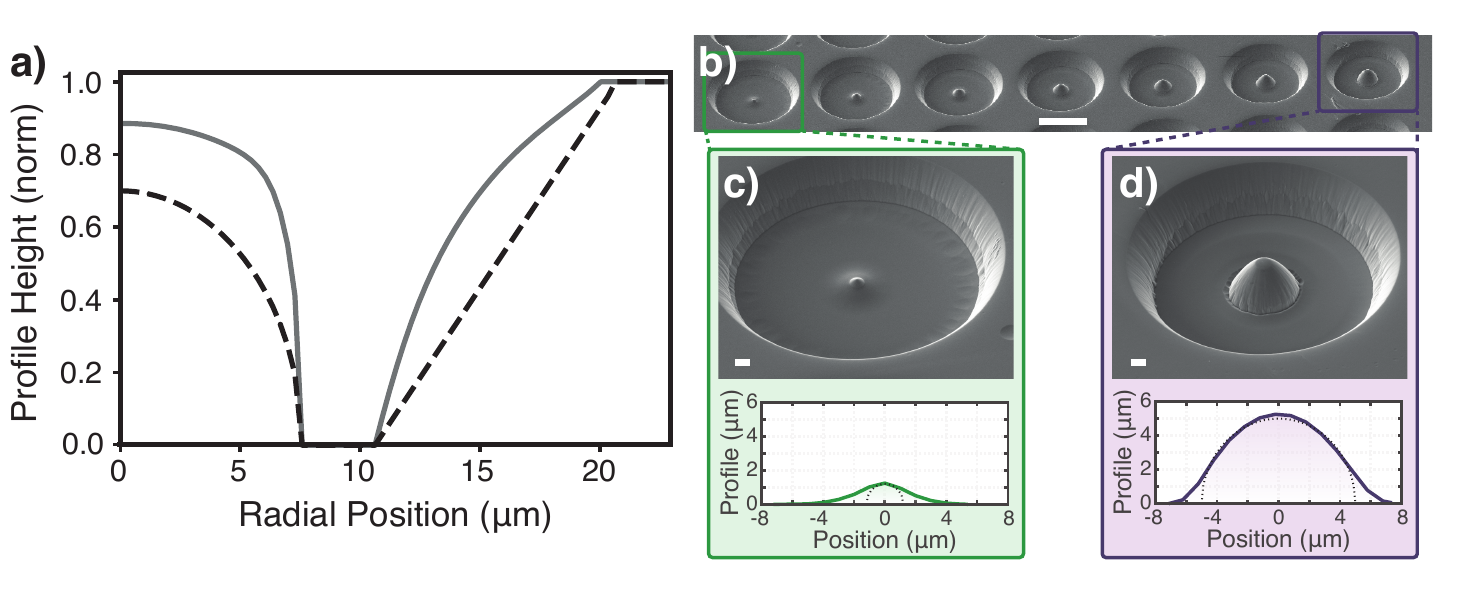}
    \caption{Analytical proximity effect correction of lens shapes~\cite{pavkovich_proximity_1986}. (a) Profile before (dashed) and after (solid) applying proximity correction with a strength of $\eta = 2$. Note that the correction changes the lens height, but does not affect the lens radius, so that a bias must also be applied. (b) SEM showing a progression of lenses (Scale bar = 20~$\mu$m) from no proximity correction (left,(c)) to a maximum strength of $\eta = 6$ (right, (d)). A hemisphere with maximum radius of 1.2~$\mu$m can be fit inside the uncorrected lens, shown by the dotted line in the profile of (c), while the strongest correction shows good agreement with a hemispherical profile of radius 5~$\mu$m, shown by the dotted line in the profile of (d). Scale bars for (c),(d) = 2~$\mu$m.}
    \label{Fig3}
\end{figure*}

\subsection{Proximity Correction}
\label{sec:proximity}

In electron-beam lithography (EBL), the random trajectories of individual electrons in the beam can lead to the exposure of sections of the resist tens of micrometers away from the beam entry point, affecting the shape of resulting resist patterns.  Proximity effect correction (PEC) is therefore routinely applied to EBL patterns to account for this and recover the desired shape. The characteristic distribution of the dose experienced from a point exposure of the resist is captured by a point-spread function (PSF), which is experimentally obtained for every combination of resist and substrate.

Direct laser writing lithography suffers similarly from proximity effects due to the beam profile and stray photons, resulting in a double-gaussian PSF \cite{du_precompensation_2000}, though these tend to be of much shorter range than experienced in EBL. However, the effects are still significant at micrometer scales, as for the lenses in this work, and so a PEC process was developed to correct for distortions in the lens shape. We use the two-gaussian profile of Du, et al.~\cite{du_precompensation_2000} as a framework for our PSF, viz:

\begin{equation}
  PSF(r) = \frac{E_0}{\pi(1+\delta)} \left[ \frac{1}{\alpha^2} e^{-\frac{r^2}{\alpha^2}} - \frac{\delta}{\beta^2} e^{-\frac{r^2}{\beta^2}}   \right]  \; , \label{Eq:PSF}
\end{equation}

where $r$ denotes the radial distance from the center of the Gaussian beam, $E_0$ is the intensity at the center of the beam, $\alpha$ is related to the width of the beam and $\beta$ and $\delta$ are experimentally-determined parameters linked to the spread and strength of scattering in the substrate, respectively. Physically, the first term in the bracket therefore describes the absorbed energy distribution due to direct illumination of the beam, and the second defines the loss caused by surface scattering and substrate absorption~\cite{du_precompensation_2000}. 

PEC is carried out following the approach of Pavkovich~\cite{pavkovich_proximity_1986}, where the required dose at a position $x$ ($D(x)$) given a desired pattern $P(x)$ and global dose $E_0$ is:

\begin{equation}
    D(x) = \frac{(E_0/K)P(x)}{\frac{1}{1+\eta} + \frac{\eta}{1+\eta}(P(x)*PSF(x))} \; , \label{Eq:PEC}
\end{equation}

where $K$ is a conversion factor between incident light intensity and energy deposition in the resist, $\eta$ is  an experimental parameter related to the strength of the correction, physically identified as the fraction of back-propagating light and so the strength of the proximity effect, and $P(x)*PSF(x)$ is the convolution of the desired pattern and PSF. 

By substituting Eq.~\ref{Eq:PSF} and the desired lens profile (Fig.~\ref{Fig3}(a), dashed black line) into Eq.~\ref{Eq:PEC}, the corrected dose profile given a set of experimental parameters $\alpha,\beta, \delta$ and $\eta$ can be obtained (Fig.~\ref{Fig3}(a), solid gray line). As can be observed in Figure~\ref{Fig3}(a), the overall effect of the proximity effect correction for this profile is to increase targeted resist height (reduce exposure dose) in the regions of partial exposure, exaggerating convex profile features. This is intuitively consistent with shapes obtained without correction, where hemispherical target profiles yielded conical lens shapes (see e.g. profiles in Fig~\ref{Fig2}(b)), so that correction could be expected to involve an increase in the curvature of the target profile.  

While optimising these values is a highly multivariate problem, the physical interpretation of each parameter provides an intuition for how changing its value will affect the final shape. Through a series of iterations and sweeps of the correction parameters, it was possible to achieve hemispherical lens shapes to a high level of fidelity. In Figure \ref{Fig3}(b), a series of lenses with increasing correction strength (increasing $\eta$) from left to right is shown. While the lowest correction strength ($\eta = 0$, reducing Eq.~\ref{Eq:PEC} to $D(x) = (E_0/K)P(x)$) shows low shape fidelity to the desired 5~$\mu$m hemisphere, the strongest correction ($\eta = 6$) produced a profile which only meaningfully deviates from a hemispherical shape at its corners, possibly due to the radius of curvature of the profilometer tip used. This clearly demonstrates the importance of proximity effect correction for laser writing of micrometer-scale profiles with large aspect ratios.

\section{Results}

\begin{figure} [ht]
    \centering
    \includegraphics{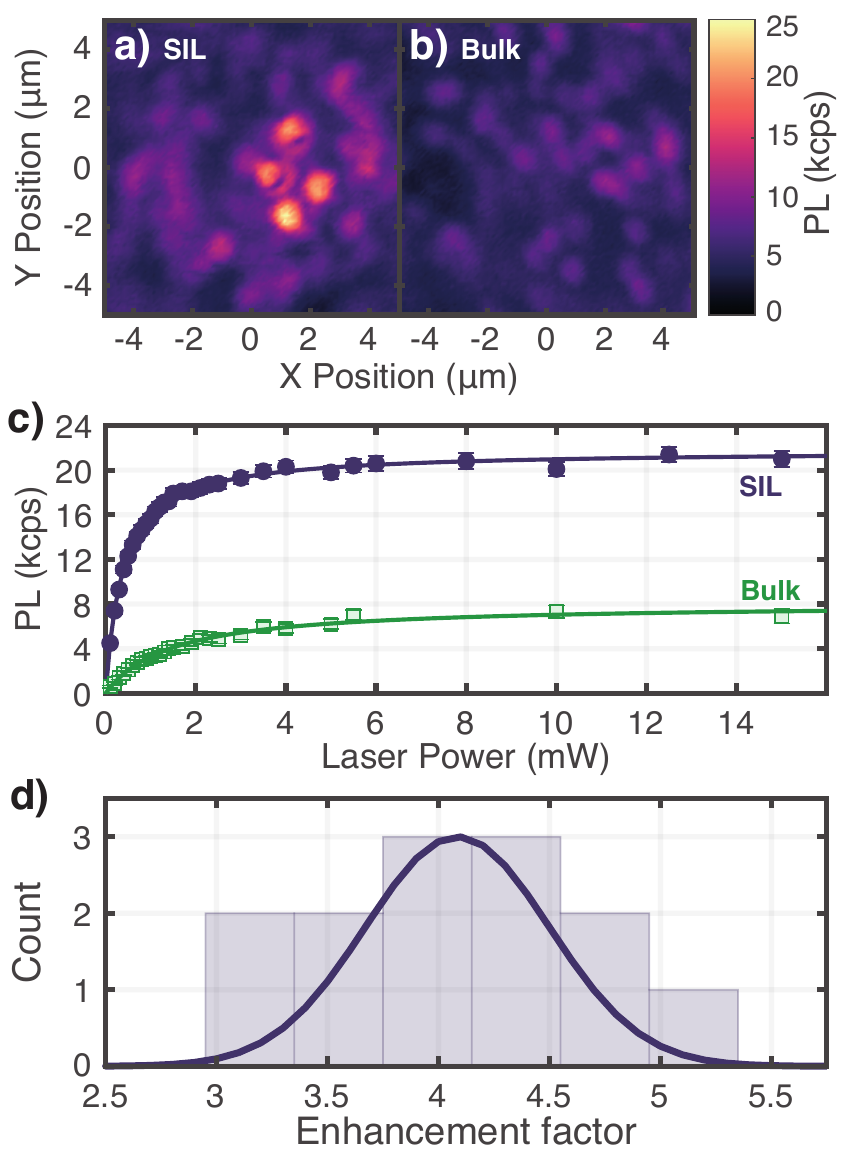}
    \caption{Experimental results of lens performance. Photoluminescence maps taken at a laser power of 3~mW of V\textsubscript{Si} emitters beneath a lens (a) and within the substrate bulk without patterned structures (b). The laser background at this power is approximately 4000 counts per second. (c) Background-subtracted power saturation measurements of single emitters in a lens (purple circles) and in the bulk (green squares), showing corresponding enhancement of the collected photoluminescence. (d) Histogram of enhancement factors observed for 13 different single V\textsubscript{Si} emitters. The enhancement factor was measured at a laser power of 3~mW and calculated by taking the ratio of the background-corrected emitter brightness with the mean background-corrected brightness of 11 emitters in the bulk.}
    \label{Fig4}
\end{figure}

We benchmark the performance of the SILs by studying the enhancement of optical collection efficiency from silicon vacancy (V$_{\text{Si}}$) centers, one of the leading systems for spin-based quantum technology in SiC~\cite{castelletto_silicon_2022, widmann_coherent_2015, ivady_identification_2017}. In order to generate V$_{\text{Si}}$ centers in the SiC material, the sample was irradiated with high-energy electrons after lens fabrication. A 10 MeV linear electron accelerator (MB10-30MP; Mevex Corp., Canada) was used for the irradiation process, with irradiation parameters: beam current = 125~mA, pulse length = 8~$\mu$s, pulse repetition rate = 460~Hz. A 10 mm thick aluminium plate in front of the sample was used to reduce the electron energy to about 4-5 MeV. The dose was determined by means of a 5 mm thick graphite calorimeter and the electron fluence then re-calculated taking a value of $\sim$2 MeV cm$^2$ g$^{-1}$ into account for the mass collision stopping power of electrons in the energy range of  4-10 MeV. Thus a the applied dose of 3.6 kGy corresponds to an electron fluence of $\sim1\times 10^{13}$ electrons per cm$^2$. 
 
Photoluminescence from these emitters was measured using a room-temperature optical confocal microscope setup, as discussed in detail in Appendix \ref{app:Setup}. V$_{\text{Si}}$ centers in SiC feature a dipole oriented along the $c$-axis, i.e. at roughly 4 degrees from the normal to the sample surface. The performance of the lenses for enhancing the collected light from single V$_{\text{Si}}$ centers was quantified through comparison of single-emitter optical properties beneath lenses (SIL emitters, see Fig. \ref{Fig4}(a)) and in regions without photonic structures (bulk emitters, see Fig. \ref{Fig4}(b)). 

In the photoluminescence maps shown in Figure \ref{Fig4}(a) and (b), the enhanced brightness of SIL emitters can be clearly ascertained. These maps were taken at a laser power of 3~mW, where both SIL and bulk emitters have reached the saturation regime (see Fig. \ref{Fig4}(d)), to ensure that the enhancement is primarily due to the enhancement of collected light, and not the focusing effect of the lens on the excitation light, which produces a smaller beam focal spot and so higher spot intensities for the same excitation power. In general, the photoluminescence intensity as a function of excitation power, $PL(I)$, of an optical emitter (as in Fig. \ref{Fig4}(d)) follows the equation:

\begin{equation} [h]
    PL(I) = PL_{\text{sat}} \frac{1}{1+\frac{I}{I_{\text{sat}}}} \; ,
\end{equation}

where $PL_{\text{sat}}$ is the photoluminescence measured at saturation and $I_{\text{sat}}$ is the excitation power when the emitter is emitting at half its saturation intensity. The first effect of the SIL, i.e. enhancement of light collection, primarily affects the saturation power of the emitter, through which the enhancement factor of the lens is determined. In the case of the emitters measured for Fig. \ref{Fig4}(d), these values for the SIL and bulk emitter are $PL_{\text{sat,SIL}} = 21.8$~kcps and $PL_{\text{sat,bulk}} =8.1$~kcps respectively, leading to an enhancement factor of 2.7 (Note that as this bulk emitter is brighter than average, this value is underestimated compared to the statistics in Fig. \ref{Fig4}(d)).  

The second effect of the SIL, the reduction of the beam spot and correspondingly increased intensity, is conversely observed as a reduction in the half-saturation laser power. For the SIL and bulk emitters measured in Fig. \ref{Fig4}(c), the value of this parameter is  $I_{\text{sat,SIL}} =0.38$~mW and $P_{\text{sat,bulk}} =1.44$~mW respectively, reading to an intensification factor of 3.8. 

In the sample measured for this experiment, the V$_{\text{Si}}$ centers are created at random locations, and variation in the enhancement factor due both to the location of SIL emitters relative to lenses and of natural distributions in brightness between different emitters can be expected. To quantify this variation, the photoluminescence at saturation of 10 bulk emitters and 13 SIL emitters (across 7 different SILs) was measured. The enhancement factor of each SIL emitter was quantified by comparing it to the average brightness of all of the bulk emitters, and the statistics for the resulting values was fit to a normal distribution, which yielded a value for the enhancement factor of $4.1 \pm 0.6$. A histogram of the data along with best fit are plotted in Fig. \ref{Fig4}(d).

It is worth noting here that lens performance can be anticipated to be similar for emitters in other materials, when changes in refractive index and dipole orientation are accounted for. Further, as the lens profile can be modified almost arbitrarily with the grayscale hardmask technique presented in this work, improved lens shapes can not only yield better performance than attained here but can be specifically optimized for other materials and emitters if desired.

\section {Conclusion and Outlook}
We have demonstrated a scalable process to fabricate hemispherical SILs using grayscale lithography, demonstrating an enhancement in optical collection efficiency of single emitters of about 4. Our approach is much faster than FIB and more flexible, in terms of shape control, than other lithographic approaches such as reflow lithography~\cite{liu_control_2016, sardi_scalable_2020}. We analyzed the SILs performance on randomly located emitters, not centered on the lens: we expect much improved performance from SILs registered to the emitter position~\cite{marseglia_nanofabricated_2011,bernien_control_2014}.

While we have focused here on hemispherical SILs, our grayscale hard-mask lithography protocol is very flexible and can enable the precise fabrication of more complex shapes. Future work will enable optimisation of the SIL shape (allowing e.g. elliptical~\cite{bogucki_ultra-long-working-distance_2020, schell_numerical_2014} or conical~\cite{sardi_scalable_2020} profiles) and adjacent structures, such as trenches, to re-direct light into lower numerical aperture collection optics.

Our technique could further be extended to different materials, such as diamond for example, provided one can achieve a sufficient balance of etching selectivity using the SiO$_2$ mask.

An appealing feature of silicon carbide is the integration of photonic structures with microelectronics, enabling the possibility to control the charge state of the emitter~\cite{widmann_electrical_2019}, tune the emission wavelength and deplete charges in the environment to minimize the spectral diffusion of the spin-selective atomic-like transitions used for spin-photon interfacing~\cite{anderson_electrical_2019}. As the grayscale hard-mask method is scalable and compatible with standard photolithgraphy and CMOS processes, it is possible to integrate the protocol with electronics workflows, allowing the co-location of photonic and electronic elements on-chip and enabling greater flexibility for hybrid photonic-electronic integrated circuits~\cite{elshaari_hybrid_2020, yoo_hybrid_2021}.

\begin{acknowledgments}

We thank Fiammetta Sardi, Matthias Widmann and Florian Kaiser for helpful discussions. We are  grateful to Daniel Andr\'es Penares and Mauro Brotons i Gisbert for their assistance with building the experimental characterization setup. We thank Dominique Colle of Heidelberg Instruments for fruitful discussions on grayscale dose calibration. 

This work is funded by the Engineering and Physical Sciences Research Council (EP/S000550/1), the Leverhulme Trust (RPG-2019-388) and the European Commission (QuanTELCO, grant agreement No 862721). C. Bekker is supported by a Royal Academy of Engineering Research Fellowship (RF2122-21-129).
\end{acknowledgments}

\bibliography{mainbib}

\begin{thebibliography}{65}%
\makeatletter
\providecommand \@ifxundefined [1]{%
 \@ifx{#1\undefined}
}%
\providecommand \@ifnum [1]{%
 \ifnum #1\expandafter \@firstoftwo
 \else \expandafter \@secondoftwo
 \fi
}%
\providecommand \@ifx [1]{%
 \ifx #1\expandafter \@firstoftwo
 \else \expandafter \@secondoftwo
 \fi
}%
\providecommand \natexlab [1]{#1}%
\providecommand \enquote  [1]{``#1''}%
\providecommand \bibnamefont  [1]{#1}%
\providecommand \bibfnamefont [1]{#1}%
\providecommand \citenamefont [1]{#1}%
\providecommand \href@noop [0]{\@secondoftwo}%
\providecommand \href [0]{\begingroup \@sanitize@url \@href}%
\providecommand \@href[1]{\@@startlink{#1}\@@href}%
\providecommand \@@href[1]{\endgroup#1\@@endlink}%
\providecommand \@sanitize@url [0]{\catcode `\\12\catcode `\$12\catcode
  `\&12\catcode `\#12\catcode `\^12\catcode `\_12\catcode `\%12\relax}%
\providecommand \@@startlink[1]{}%
\providecommand \@@endlink[0]{}%
\providecommand \url  [0]{\begingroup\@sanitize@url \@url }%
\providecommand \@url [1]{\endgroup\@href {#1}{\urlprefix }}%
\providecommand \urlprefix  [0]{URL }%
\providecommand \Eprint [0]{\href }%
\providecommand \doibase [0]{http://dx.doi.org/}%
\providecommand \selectlanguage [0]{\@gobble}%
\providecommand \bibinfo  [0]{\@secondoftwo}%
\providecommand \bibfield  [0]{\@secondoftwo}%
\providecommand \translation [1]{[#1]}%
\providecommand \BibitemOpen [0]{}%
\providecommand \bibitemStop [0]{}%
\providecommand \bibitemNoStop [0]{.\EOS\space}%
\providecommand \EOS [0]{\spacefactor3000\relax}%
\providecommand \BibitemShut  [1]{\csname bibitem#1\endcsname}%
\let\auto@bib@innerbib\@empty
\bibitem [{\citenamefont {Cai}\ \emph {et~al.}(2021)\citenamefont {Cai},
  \citenamefont {Sun}, \citenamefont {Chu}, \citenamefont {Yang}, \citenamefont
  {Yu},\ and\ \citenamefont {Liu}}]{cai_microlenses_2021}%
  \BibitemOpen
  \bibfield  {author} {\bibinfo {author} {\bibfnamefont {S.}~\bibnamefont
  {Cai}}, \bibinfo {author} {\bibfnamefont {Y.}~\bibnamefont {Sun}}, \bibinfo
  {author} {\bibfnamefont {H.}~\bibnamefont {Chu}}, \bibinfo {author}
  {\bibfnamefont {W.}~\bibnamefont {Yang}}, \bibinfo {author} {\bibfnamefont
  {H.}~\bibnamefont {Yu}}, \ and\ \bibinfo {author} {\bibfnamefont
  {L.}~\bibnamefont {Liu}},\ }\bibfield  {title} {\enquote {\bibinfo {title}
  {Microlenses arrays: {Fabrication}, materials, and applications},}\ }\href
  {\doibase 10.1002/jemt.23818} {\bibfield  {journal} {\bibinfo  {journal}
  {Microscopy Research and Technique}\ }\textbf {\bibinfo {volume} {84}},\
  \bibinfo {pages} {2784--2806} (\bibinfo {year} {2021})}\BibitemShut {NoStop}%
\bibitem [{\citenamefont {Yuan}\ \emph {et~al.}(2018)\citenamefont {Yuan},
  \citenamefont {Li}, \citenamefont {Lee},\ and\ \citenamefont
  {Chan}}]{yuan_fabrication_2018}%
  \BibitemOpen
  \bibfield  {author} {\bibinfo {author} {\bibfnamefont {W.}~\bibnamefont
  {Yuan}}, \bibinfo {author} {\bibfnamefont {L.-H.}\ \bibnamefont {Li}},
  \bibinfo {author} {\bibfnamefont {W.-B.}\ \bibnamefont {Lee}}, \ and\
  \bibinfo {author} {\bibfnamefont {C.-Y.}\ \bibnamefont {Chan}},\ }\bibfield
  {title} {\enquote {\bibinfo {title} {Fabrication of {Microlens} {Array} and
  {Its} {Application}: {A} {Review}},}\ }\href {\doibase
  10.1186/s10033-018-0204-y} {\bibfield  {journal} {\bibinfo  {journal}
  {Chinese Journal of Mechanical Engineering}\ }\textbf {\bibinfo {volume}
  {31}},\ \bibinfo {pages} {16} (\bibinfo {year} {2018})}\BibitemShut {NoStop}%
\bibitem [{\citenamefont {Nussbaum}\ \emph {et~al.}(1997)\citenamefont
  {Nussbaum}, \citenamefont {Völkel}, \citenamefont {Herzig}, \citenamefont
  {Eisner},\ and\ \citenamefont {Haselbeck}}]{nussbaum_design_1997}%
  \BibitemOpen
  \bibfield  {author} {\bibinfo {author} {\bibfnamefont {P.}~\bibnamefont
  {Nussbaum}}, \bibinfo {author} {\bibfnamefont {R.}~\bibnamefont {Völkel}},
  \bibinfo {author} {\bibfnamefont {H.~P.}\ \bibnamefont {Herzig}}, \bibinfo
  {author} {\bibfnamefont {M.}~\bibnamefont {Eisner}}, \ and\ \bibinfo {author}
  {\bibfnamefont {S.}~\bibnamefont {Haselbeck}},\ }\bibfield  {title} {\enquote
  {\bibinfo {title} {Design, fabrication and testing of microlens arrays for
  sensors and microsystems},}\ }\href {\doibase 10.1088/0963-9659/6/6/004}
  {\bibfield  {journal} {\bibinfo  {journal} {Pure and Applied Optics: Journal
  of the European Optical Society Part A}\ }\textbf {\bibinfo {volume} {6}},\
  \bibinfo {pages} {617--636} (\bibinfo {year} {1997})}\BibitemShut {NoStop}%
\bibitem [{\citenamefont {Tan}\ \emph {et~al.}(2020)\citenamefont {Tan},
  \citenamefont {Yap}, \citenamefont {Wong}, \citenamefont {Ng}, \citenamefont
  {Ng}, \citenamefont {Grenci}, \citenamefont {Grenci}, \citenamefont {Grenci},
  \citenamefont {Danner},\ and\ \citenamefont {Danner}}]{tan_high_2020}%
  \BibitemOpen
  \bibfield  {author} {\bibinfo {author} {\bibfnamefont {S.~L.}\ \bibnamefont
  {Tan}}, \bibinfo {author} {\bibfnamefont {Y.~K.}\ \bibnamefont {Yap}},
  \bibinfo {author} {\bibfnamefont {J.~J.}\ \bibnamefont {Wong}}, \bibinfo
  {author} {\bibfnamefont {J.~D.}\ \bibnamefont {Ng}}, \bibinfo {author}
  {\bibfnamefont {J.~D.}\ \bibnamefont {Ng}}, \bibinfo {author} {\bibfnamefont
  {G.}~\bibnamefont {Grenci}}, \bibinfo {author} {\bibfnamefont
  {G.}~\bibnamefont {Grenci}}, \bibinfo {author} {\bibfnamefont
  {G.}~\bibnamefont {Grenci}}, \bibinfo {author} {\bibfnamefont {A.~J.}\
  \bibnamefont {Danner}}, \ and\ \bibinfo {author} {\bibfnamefont {A.~J.}\
  \bibnamefont {Danner}},\ }\bibfield  {title} {\enquote {\bibinfo {title}
  {High pulsed power {VCSEL} arrays with polymer microlenses formed by
  photoacid diffusion},}\ }\href {\doibase 10.1364/OE.392660} {\bibfield
  {journal} {\bibinfo  {journal} {Optics Express}\ }\textbf {\bibinfo {volume}
  {28}},\ \bibinfo {pages} {20095--20105} (\bibinfo {year} {2020})}\BibitemShut
  {NoStop}%
\bibitem [{\citenamefont {Wang}\ \emph {et~al.}(2010)\citenamefont {Wang},
  \citenamefont {Ning}, \citenamefont {Zhang}, \citenamefont {Shi},
  \citenamefont {Zhang}, \citenamefont {Zhang}, \citenamefont {Wang},
  \citenamefont {Liu}, \citenamefont {Hu}, \citenamefont {Cong}, \citenamefont
  {Qin}, \citenamefont {Liu},\ and\ \citenamefont {Wang}}]{wang_high_2010}%
  \BibitemOpen
  \bibfield  {author} {\bibinfo {author} {\bibfnamefont {Z.}~\bibnamefont
  {Wang}}, \bibinfo {author} {\bibfnamefont {Y.}~\bibnamefont {Ning}}, \bibinfo
  {author} {\bibfnamefont {Y.}~\bibnamefont {Zhang}}, \bibinfo {author}
  {\bibfnamefont {J.}~\bibnamefont {Shi}}, \bibinfo {author} {\bibfnamefont
  {X.}~\bibnamefont {Zhang}}, \bibinfo {author} {\bibfnamefont
  {L.}~\bibnamefont {Zhang}}, \bibinfo {author} {\bibfnamefont
  {W.}~\bibnamefont {Wang}}, \bibinfo {author} {\bibfnamefont {D.}~\bibnamefont
  {Liu}}, \bibinfo {author} {\bibfnamefont {Y.}~\bibnamefont {Hu}}, \bibinfo
  {author} {\bibfnamefont {H.}~\bibnamefont {Cong}}, \bibinfo {author}
  {\bibfnamefont {L.}~\bibnamefont {Qin}}, \bibinfo {author} {\bibfnamefont
  {Y.}~\bibnamefont {Liu}}, \ and\ \bibinfo {author} {\bibfnamefont
  {L.}~\bibnamefont {Wang}},\ }\bibfield  {title} {\enquote {\bibinfo {title}
  {High power and good beam quality of two-dimensional {VCSEL} array with
  integrated {GaAs} microlens array},}\ }\href {\doibase 10.1364/OE.18.023900}
  {\bibfield  {journal} {\bibinfo  {journal} {Optics Express}\ }\textbf
  {\bibinfo {volume} {18}},\ \bibinfo {pages} {23900--23905} (\bibinfo {year}
  {2010})}\BibitemShut {NoStop}%
\bibitem [{\citenamefont {Gschrey}\ \emph {et~al.}(2015)\citenamefont
  {Gschrey}, \citenamefont {Thoma}, \citenamefont {Schnauber}, \citenamefont
  {Seifried}, \citenamefont {Schmidt}, \citenamefont {Wohlfeil}, \citenamefont
  {Krüger}, \citenamefont {Schulze}, \citenamefont {Heindel}, \citenamefont
  {Burger}, \citenamefont {Schmidt}, \citenamefont {Strittmatter},
  \citenamefont {Rodt},\ and\ \citenamefont
  {Reitzenstein}}]{gschrey_highly_2015}%
  \BibitemOpen
  \bibfield  {author} {\bibinfo {author} {\bibfnamefont {M.}~\bibnamefont
  {Gschrey}}, \bibinfo {author} {\bibfnamefont {A.}~\bibnamefont {Thoma}},
  \bibinfo {author} {\bibfnamefont {P.}~\bibnamefont {Schnauber}}, \bibinfo
  {author} {\bibfnamefont {M.}~\bibnamefont {Seifried}}, \bibinfo {author}
  {\bibfnamefont {R.}~\bibnamefont {Schmidt}}, \bibinfo {author} {\bibfnamefont
  {B.}~\bibnamefont {Wohlfeil}}, \bibinfo {author} {\bibfnamefont
  {L.}~\bibnamefont {Krüger}}, \bibinfo {author} {\bibfnamefont {J.-H.}\
  \bibnamefont {Schulze}}, \bibinfo {author} {\bibfnamefont {T.}~\bibnamefont
  {Heindel}}, \bibinfo {author} {\bibfnamefont {S.}~\bibnamefont {Burger}},
  \bibinfo {author} {\bibfnamefont {F.}~\bibnamefont {Schmidt}}, \bibinfo
  {author} {\bibfnamefont {A.}~\bibnamefont {Strittmatter}}, \bibinfo {author}
  {\bibfnamefont {S.}~\bibnamefont {Rodt}}, \ and\ \bibinfo {author}
  {\bibfnamefont {S.}~\bibnamefont {Reitzenstein}},\ }\bibfield  {title}
  {\enquote {\bibinfo {title} {Highly indistinguishable photons from
  deterministic quantum-dot microlenses utilizing three-dimensional in situ
  electron-beam lithography},}\ }\href {\doibase 10.1038/ncomms8662} {\bibfield
   {journal} {\bibinfo  {journal} {Nature Communications}\ }\textbf {\bibinfo
  {volume} {6}},\ \bibinfo {pages} {7662} (\bibinfo {year} {2015})}\BibitemShut
  {NoStop}%
\bibitem [{\citenamefont {Bogucki}\ \emph {et~al.}(2020)\citenamefont
  {Bogucki}, \citenamefont {Zinkiewicz}, \citenamefont {Grzeszczyk},
  \citenamefont {Pacuski}, \citenamefont {Nogajewski}, \citenamefont
  {Kazimierczuk}, \citenamefont {Rodek}, \citenamefont {Suffczyński},
  \citenamefont {Watanabe}, \citenamefont {Taniguchi}, \citenamefont
  {Wasylczyk}, \citenamefont {Potemski},\ and\ \citenamefont
  {Kossacki}}]{bogucki_ultra-long-working-distance_2020}%
  \BibitemOpen
  \bibfield  {author} {\bibinfo {author} {\bibfnamefont {A.}~\bibnamefont
  {Bogucki}}, \bibinfo {author} {\bibfnamefont {{\L}.}~\bibnamefont
  {Zinkiewicz}}, \bibinfo {author} {\bibfnamefont {M.}~\bibnamefont
  {Grzeszczyk}}, \bibinfo {author} {\bibfnamefont {W.}~\bibnamefont {Pacuski}},
  \bibinfo {author} {\bibfnamefont {K.}~\bibnamefont {Nogajewski}}, \bibinfo
  {author} {\bibfnamefont {T.}~\bibnamefont {Kazimierczuk}}, \bibinfo {author}
  {\bibfnamefont {A.}~\bibnamefont {Rodek}}, \bibinfo {author} {\bibfnamefont
  {J.}~\bibnamefont {Suffczyński}}, \bibinfo {author} {\bibfnamefont
  {K.}~\bibnamefont {Watanabe}}, \bibinfo {author} {\bibfnamefont
  {T.}~\bibnamefont {Taniguchi}}, \bibinfo {author} {\bibfnamefont
  {P.}~\bibnamefont {Wasylczyk}}, \bibinfo {author} {\bibfnamefont
  {M.}~\bibnamefont {Potemski}}, \ and\ \bibinfo {author} {\bibfnamefont
  {P.}~\bibnamefont {Kossacki}},\ }\bibfield  {title} {\enquote {\bibinfo
  {title} {Ultra-long-working-distance spectroscopy of single nanostructures
  with aspherical solid immersion microlenses},}\ }\href {\doibase
  10.1038/s41377-020-0284-1} {\bibfield  {journal} {\bibinfo  {journal} {Light:
  Science \& Applications}\ }\textbf {\bibinfo {volume} {9}},\ \bibinfo {pages}
  {48} (\bibinfo {year} {2020})}\BibitemShut {NoStop}%
\bibitem [{\citenamefont {Preuß}\ \emph {et~al.}(2022)\citenamefont {Preuß},
  \citenamefont {Gehring}, \citenamefont {Schmidt}, \citenamefont {Jin},
  \citenamefont {Wendland}, \citenamefont {Kern}, \citenamefont {Pernice},
  \citenamefont {de~Vasconcellos},\ and\ \citenamefont
  {Bratschitsch}}]{preus_low-divergence_2022}%
  \BibitemOpen
  \bibfield  {author} {\bibinfo {author} {\bibfnamefont {J.~A.}\ \bibnamefont
  {Preuß}}, \bibinfo {author} {\bibfnamefont {H.}~\bibnamefont {Gehring}},
  \bibinfo {author} {\bibfnamefont {R.}~\bibnamefont {Schmidt}}, \bibinfo
  {author} {\bibfnamefont {L.}~\bibnamefont {Jin}}, \bibinfo {author}
  {\bibfnamefont {D.}~\bibnamefont {Wendland}}, \bibinfo {author}
  {\bibfnamefont {J.}~\bibnamefont {Kern}}, \bibinfo {author} {\bibfnamefont
  {W.~H.~P.}\ \bibnamefont {Pernice}}, \bibinfo {author} {\bibfnamefont
  {S.~M.}\ \bibnamefont {de~Vasconcellos}}, \ and\ \bibinfo {author}
  {\bibfnamefont {R.}~\bibnamefont {Bratschitsch}},\ }\bibfield  {title}
  {\enquote {\bibinfo {title} {Low-{Divergence} {hBN} {Single}-{Photon}
  {Source} with a {3D}-{Printed} {Low}-{Fluorescence} {Elliptical} {Polymer}
  {Microlens}},}\ }\href {\doibase 10.1021/acs.nanolett.2c03001} {\bibfield
  {journal} {\bibinfo  {journal} {Nano Letters}\ } (\bibinfo {year} {2022}),\
  10.1021/acs.nanolett.2c03001}\BibitemShut {NoStop}%
\bibitem [{\citenamefont {Li}\ \emph {et~al.}(2021)\citenamefont {Li},
  \citenamefont {Shang}, \citenamefont {Chen}, \citenamefont {Su},
  \citenamefont {Hao}, \citenamefont {Liu}, \citenamefont {Zhang},
  \citenamefont {Ni},\ and\ \citenamefont {Niu}}]{li_wet-etched_2021}%
  \BibitemOpen
  \bibfield  {author} {\bibinfo {author} {\bibfnamefont {S.}~\bibnamefont
  {Li}}, \bibinfo {author} {\bibfnamefont {X.}~\bibnamefont {Shang}}, \bibinfo
  {author} {\bibfnamefont {Y.}~\bibnamefont {Chen}}, \bibinfo {author}
  {\bibfnamefont {X.}~\bibnamefont {Su}}, \bibinfo {author} {\bibfnamefont
  {H.}~\bibnamefont {Hao}}, \bibinfo {author} {\bibfnamefont {H.}~\bibnamefont
  {Liu}}, \bibinfo {author} {\bibfnamefont {Y.}~\bibnamefont {Zhang}}, \bibinfo
  {author} {\bibfnamefont {H.}~\bibnamefont {Ni}}, \ and\ \bibinfo {author}
  {\bibfnamefont {Z.}~\bibnamefont {Niu}},\ }\bibfield  {title} {\enquote
  {\bibinfo {title} {Wet-{Etched} {Microlens} {Array} for 200 nm {Spatial}
  {Isolation} of {Epitaxial} {Single} {QDs} and 80 nm {Broadband} {Enhancement}
  of {Their} {Quantum} {Light} {Extraction}},}\ }\href {\doibase
  10.3390/nano11051136} {\bibfield  {journal} {\bibinfo  {journal}
  {Nanomaterials}\ }\textbf {\bibinfo {volume} {11}},\ \bibinfo {pages} {1136}
  (\bibinfo {year} {2021})}\BibitemShut {NoStop}%
\bibitem [{\citenamefont {Schlingloff}, \citenamefont {Kiel},\ and\
  \citenamefont {Schober}(1998)}]{schlingloff_microlenses_1998}%
  \BibitemOpen
  \bibfield  {author} {\bibinfo {author} {\bibfnamefont {G.}~\bibnamefont
  {Schlingloff}}, \bibinfo {author} {\bibfnamefont {H.-J.}\ \bibnamefont
  {Kiel}}, \ and\ \bibinfo {author} {\bibfnamefont {A.}~\bibnamefont
  {Schober}},\ }\bibfield  {title} {\enquote {\bibinfo {title} {Microlenses as
  amplification for {CCD}-based detection devices for screening applications in
  biology, biochemistry, and chemistry},}\ }\href {\doibase
  10.1364/AO.37.001930} {\bibfield  {journal} {\bibinfo  {journal} {Applied
  Optics}\ }\textbf {\bibinfo {volume} {37}},\ \bibinfo {pages} {1930--1934}
  (\bibinfo {year} {1998})}\BibitemShut {NoStop}%
\bibitem [{\citenamefont {Ke}\ \emph {et~al.}(2005)\citenamefont {Ke},
  \citenamefont {Yi}, \citenamefont {Xu},\ and\ \citenamefont
  {Lai}}]{ke_monolithic_2005}%
  \BibitemOpen
  \bibfield  {author} {\bibinfo {author} {\bibfnamefont {C.}~\bibnamefont
  {Ke}}, \bibinfo {author} {\bibfnamefont {X.}~\bibnamefont {Yi}}, \bibinfo
  {author} {\bibfnamefont {Z.}~\bibnamefont {Xu}}, \ and\ \bibinfo {author}
  {\bibfnamefont {J.}~\bibnamefont {Lai}},\ }\bibfield  {title} {\enquote
  {\bibinfo {title} {Monolithic integration technology between microlens arrays
  and infrared charge coupled devices},}\ }\href {\doibase
  10.1016/j.optlastec.2004.04.002} {\bibfield  {journal} {\bibinfo  {journal}
  {Optics \& Laser Technology}\ }\textbf {\bibinfo {volume} {37}},\ \bibinfo
  {pages} {239--243} (\bibinfo {year} {2005})}\BibitemShut {NoStop}%
\bibitem [{\citenamefont {Tripathi}\ and\ \citenamefont
  {Chronis}(2011)}]{tripathi_doublet_2011}%
  \BibitemOpen
  \bibfield  {author} {\bibinfo {author} {\bibfnamefont {A.}~\bibnamefont
  {Tripathi}}\ and\ \bibinfo {author} {\bibfnamefont {N.}~\bibnamefont
  {Chronis}},\ }\bibfield  {title} {\enquote {\bibinfo {title} {A doublet
  microlens array for imaging micron-sized objects},}\ }\href {\doibase
  10.1088/0960-1317/21/10/105024} {\bibfield  {journal} {\bibinfo  {journal}
  {Journal of micromechanics and microengineering : structures, devices, and
  systems}\ }\textbf {\bibinfo {volume} {21}},\ \bibinfo {pages} {105024}
  (\bibinfo {year} {2011})}\BibitemShut {NoStop}%
\bibitem [{\citenamefont {McCormick}\ \emph {et~al.}(1992)\citenamefont
  {McCormick}, \citenamefont {Tooley}, \citenamefont {Cloonan}, \citenamefont
  {Sasian}, \citenamefont {Hinton}, \citenamefont {Mersereau},\ and\
  \citenamefont {Feldblum}}]{mccormick_optical_1992}%
  \BibitemOpen
  \bibfield  {author} {\bibinfo {author} {\bibfnamefont {F.~B.}\ \bibnamefont
  {McCormick}}, \bibinfo {author} {\bibfnamefont {F.~A.~P.}\ \bibnamefont
  {Tooley}}, \bibinfo {author} {\bibfnamefont {T.~J.}\ \bibnamefont {Cloonan}},
  \bibinfo {author} {\bibfnamefont {J.~M.}\ \bibnamefont {Sasian}}, \bibinfo
  {author} {\bibfnamefont {H.~S.}\ \bibnamefont {Hinton}}, \bibinfo {author}
  {\bibfnamefont {K.~O.}\ \bibnamefont {Mersereau}}, \ and\ \bibinfo {author}
  {\bibfnamefont {A.~Y.}\ \bibnamefont {Feldblum}},\ }\bibfield  {title}
  {\enquote {\bibinfo {title} {Optical interconnections using microlens
  arrays},}\ }\href {\doibase 10.1007/BF00619512} {\bibfield  {journal}
  {\bibinfo  {journal} {Optical and Quantum Electronics}\ }\textbf {\bibinfo
  {volume} {24}},\ \bibinfo {pages} {S465--S477} (\bibinfo {year}
  {1992})}\BibitemShut {NoStop}%
\bibitem [{\citenamefont {Tang}\ \emph {et~al.}(1996)\citenamefont {Tang},
  \citenamefont {Li}, \citenamefont {Li}, \citenamefont {Zhou},\ and\
  \citenamefont {Chen}}]{tang_holographic_1996}%
  \BibitemOpen
  \bibfield  {author} {\bibinfo {author} {\bibfnamefont {S.}~\bibnamefont
  {Tang}}, \bibinfo {author} {\bibfnamefont {T.}~\bibnamefont {Li}}, \bibinfo
  {author} {\bibfnamefont {F.}~\bibnamefont {Li}}, \bibinfo {author}
  {\bibfnamefont {C.}~\bibnamefont {Zhou}}, \ and\ \bibinfo {author}
  {\bibfnamefont {R.}~\bibnamefont {Chen}},\ }\bibfield  {title} {\enquote
  {\bibinfo {title} {A holographic waveguide microlens array for surface-normal
  optical interconnects},}\ }\href {\doibase 10.1109/68.541562} {\bibfield
  {journal} {\bibinfo  {journal} {IEEE Photonics Technology Letters}\ }\textbf
  {\bibinfo {volume} {8}},\ \bibinfo {pages} {1498--1500} (\bibinfo {year}
  {1996})}\BibitemShut {NoStop}%
\bibitem [{\citenamefont {Ballen}\ and\ \citenamefont
  {Leger}(2000)}]{ballen_off-axis_2000}%
  \BibitemOpen
  \bibfield  {author} {\bibinfo {author} {\bibfnamefont {T.~A.}\ \bibnamefont
  {Ballen}}\ and\ \bibinfo {author} {\bibfnamefont {J.~R.}\ \bibnamefont
  {Leger}},\ }\bibfield  {title} {\enquote {\bibinfo {title} {Off-axis
  refractive mass-transported gallium-phosphide microlens array for the
  reduction of distortion in an optical interconnect system},}\ }\href
  {\doibase 10.1364/AO.39.006028} {\bibfield  {journal} {\bibinfo  {journal}
  {Applied Optics}\ }\textbf {\bibinfo {volume} {39}},\ \bibinfo {pages}
  {6028--6033} (\bibinfo {year} {2000})}\BibitemShut {NoStop}%
\bibitem [{\citenamefont {Marseglia}\ \emph {et~al.}(2011)\citenamefont
  {Marseglia}, \citenamefont {Hadden}, \citenamefont {Stanley-Clarke},
  \citenamefont {Harrison}, \citenamefont {Patton}, \citenamefont {Ho},
  \citenamefont {Naydenov}, \citenamefont {Jelezko}, \citenamefont {Meijer},
  \citenamefont {Dolan}, \citenamefont {Smith}, \citenamefont {Rarity},\ and\
  \citenamefont {O’Brien}}]{marseglia_nanofabricated_2011}%
  \BibitemOpen
  \bibfield  {author} {\bibinfo {author} {\bibfnamefont {L.}~\bibnamefont
  {Marseglia}}, \bibinfo {author} {\bibfnamefont {J.~P.}\ \bibnamefont
  {Hadden}}, \bibinfo {author} {\bibfnamefont {A.~C.}\ \bibnamefont
  {Stanley-Clarke}}, \bibinfo {author} {\bibfnamefont {J.~P.}\ \bibnamefont
  {Harrison}}, \bibinfo {author} {\bibfnamefont {B.}~\bibnamefont {Patton}},
  \bibinfo {author} {\bibfnamefont {Y.-L.~D.}\ \bibnamefont {Ho}}, \bibinfo
  {author} {\bibfnamefont {B.}~\bibnamefont {Naydenov}}, \bibinfo {author}
  {\bibfnamefont {F.}~\bibnamefont {Jelezko}}, \bibinfo {author} {\bibfnamefont
  {J.}~\bibnamefont {Meijer}}, \bibinfo {author} {\bibfnamefont {P.~R.}\
  \bibnamefont {Dolan}}, \bibinfo {author} {\bibfnamefont {J.~M.}\ \bibnamefont
  {Smith}}, \bibinfo {author} {\bibfnamefont {J.~G.}\ \bibnamefont {Rarity}}, \
  and\ \bibinfo {author} {\bibfnamefont {J.~L.}\ \bibnamefont {O’Brien}},\
  }\bibfield  {title} {\enquote {\bibinfo {title} {Nanofabricated solid
  immersion lenses registered to single emitters in diamond},}\ }\href
  {\doibase 10.1063/1.3573870} {\bibfield  {journal} {\bibinfo  {journal}
  {Applied Physics Letters}\ }\textbf {\bibinfo {volume} {98}},\ \bibinfo
  {pages} {133107} (\bibinfo {year} {2011})}\BibitemShut {NoStop}%
\bibitem [{\citenamefont {Hadden}\ \emph {et~al.}(2010)\citenamefont {Hadden},
  \citenamefont {Harrison}, \citenamefont {Stanley-Clarke}, \citenamefont
  {Marseglia}, \citenamefont {Ho}, \citenamefont {Patton}, \citenamefont
  {O’Brien},\ and\ \citenamefont {Rarity}}]{hadden_strongly_2010}%
  \BibitemOpen
  \bibfield  {author} {\bibinfo {author} {\bibfnamefont {J.~P.}\ \bibnamefont
  {Hadden}}, \bibinfo {author} {\bibfnamefont {J.~P.}\ \bibnamefont
  {Harrison}}, \bibinfo {author} {\bibfnamefont {A.~C.}\ \bibnamefont
  {Stanley-Clarke}}, \bibinfo {author} {\bibfnamefont {L.}~\bibnamefont
  {Marseglia}}, \bibinfo {author} {\bibfnamefont {Y.-L.~D.}\ \bibnamefont
  {Ho}}, \bibinfo {author} {\bibfnamefont {B.~R.}\ \bibnamefont {Patton}},
  \bibinfo {author} {\bibfnamefont {J.~L.}\ \bibnamefont {O’Brien}}, \ and\
  \bibinfo {author} {\bibfnamefont {J.~G.}\ \bibnamefont {Rarity}},\ }\bibfield
   {title} {\enquote {\bibinfo {title} {Strongly enhanced photon collection
  from diamond defect centers under microfabricated integrated solid immersion
  lenses},}\ }\href {\doibase 10.1063/1.3519847} {\bibfield  {journal}
  {\bibinfo  {journal} {Applied Physics Letters}\ }\textbf {\bibinfo {volume}
  {97}},\ \bibinfo {pages} {241901} (\bibinfo {year} {2010})}\BibitemShut
  {NoStop}%
\bibitem [{\citenamefont {Robledo}\ \emph {et~al.}(2011)\citenamefont
  {Robledo}, \citenamefont {Childress}, \citenamefont {Bernien}, \citenamefont
  {Hensen}, \citenamefont {Alkemade},\ and\ \citenamefont
  {Hanson}}]{robledo_high-fidelity_2011}%
  \BibitemOpen
  \bibfield  {author} {\bibinfo {author} {\bibfnamefont {L.}~\bibnamefont
  {Robledo}}, \bibinfo {author} {\bibfnamefont {L.}~\bibnamefont {Childress}},
  \bibinfo {author} {\bibfnamefont {H.}~\bibnamefont {Bernien}}, \bibinfo
  {author} {\bibfnamefont {B.}~\bibnamefont {Hensen}}, \bibinfo {author}
  {\bibfnamefont {P.~F.~A.}\ \bibnamefont {Alkemade}}, \ and\ \bibinfo {author}
  {\bibfnamefont {R.}~\bibnamefont {Hanson}},\ }\bibfield  {title} {\enquote
  {\bibinfo {title} {High-fidelity projective read-out of a solid-state spin
  quantum register},}\ }\href {\doibase 10.1038/nature10401} {\bibfield
  {journal} {\bibinfo  {journal} {Nature}\ }\textbf {\bibinfo {volume} {477}},\
  \bibinfo {pages} {574--578} (\bibinfo {year} {2011})}\BibitemShut {NoStop}%
\bibitem [{\citenamefont {Bernien}(2014)}]{bernien_control_2014}%
  \BibitemOpen
  \bibfield  {author} {\bibinfo {author} {\bibfnamefont {H.}~\bibnamefont
  {Bernien}},\ }\emph {\bibinfo {title} {Control, measurement and entanglement
  of remote quantum spin registers in diamond}},\ \href
  {https://repository.tudelft.nl/islandora/object/uuid%3A75130c37-edb5-4a34-ac2f-c156d377ca55}
  {Ph.D. thesis},\ \bibinfo  {school} {TUDelft} (\bibinfo {year}
  {2014})\BibitemShut {NoStop}%
\bibitem [{\citenamefont {Hensen}\ \emph {et~al.}(2015)\citenamefont {Hensen},
  \citenamefont {Bernien}, \citenamefont {Dréau}, \citenamefont {Reiserer},
  \citenamefont {Kalb}, \citenamefont {Blok}, \citenamefont {Ruitenberg},
  \citenamefont {Vermeulen}, \citenamefont {Schouten}, \citenamefont
  {Abellán}, \citenamefont {Amaya}, \citenamefont {Pruneri}, \citenamefont
  {Mitchell}, \citenamefont {Markham}, \citenamefont {Twitchen}, \citenamefont
  {Elkouss}, \citenamefont {Wehner}, \citenamefont {Taminiau},\ and\
  \citenamefont {Hanson}}]{hensen_loophole-free_2015}%
  \BibitemOpen
  \bibfield  {author} {\bibinfo {author} {\bibfnamefont {B.}~\bibnamefont
  {Hensen}}, \bibinfo {author} {\bibfnamefont {H.}~\bibnamefont {Bernien}},
  \bibinfo {author} {\bibfnamefont {A.~E.}\ \bibnamefont {Dréau}}, \bibinfo
  {author} {\bibfnamefont {A.}~\bibnamefont {Reiserer}}, \bibinfo {author}
  {\bibfnamefont {N.}~\bibnamefont {Kalb}}, \bibinfo {author} {\bibfnamefont
  {M.~S.}\ \bibnamefont {Blok}}, \bibinfo {author} {\bibfnamefont
  {J.}~\bibnamefont {Ruitenberg}}, \bibinfo {author} {\bibfnamefont {R.~F.~L.}\
  \bibnamefont {Vermeulen}}, \bibinfo {author} {\bibfnamefont {R.~N.}\
  \bibnamefont {Schouten}}, \bibinfo {author} {\bibfnamefont {C.}~\bibnamefont
  {Abellán}}, \bibinfo {author} {\bibfnamefont {W.}~\bibnamefont {Amaya}},
  \bibinfo {author} {\bibfnamefont {V.}~\bibnamefont {Pruneri}}, \bibinfo
  {author} {\bibfnamefont {M.~W.}\ \bibnamefont {Mitchell}}, \bibinfo {author}
  {\bibfnamefont {M.}~\bibnamefont {Markham}}, \bibinfo {author} {\bibfnamefont
  {D.~J.}\ \bibnamefont {Twitchen}}, \bibinfo {author} {\bibfnamefont
  {D.}~\bibnamefont {Elkouss}}, \bibinfo {author} {\bibfnamefont
  {S.}~\bibnamefont {Wehner}}, \bibinfo {author} {\bibfnamefont {T.~H.}\
  \bibnamefont {Taminiau}}, \ and\ \bibinfo {author} {\bibfnamefont
  {R.}~\bibnamefont {Hanson}},\ }\bibfield  {title} {\enquote {\bibinfo {title}
  {Loophole-free {Bell} inequality violation using electron spins separated by
  1.3 kilometres},}\ }\href {\doibase 10.1038/nature15759} {\bibfield
  {journal} {\bibinfo  {journal} {Nature}\ }\textbf {\bibinfo {volume} {526}},\
  \bibinfo {pages} {682--686} (\bibinfo {year} {2015})}\BibitemShut {NoStop}%
\bibitem [{\citenamefont {Pompili}\ \emph {et~al.}(2021)\citenamefont
  {Pompili}, \citenamefont {Hermans}, \citenamefont {Baier}, \citenamefont
  {Beukers}, \citenamefont {Humphreys}, \citenamefont {Schouten}, \citenamefont
  {Vermeulen}, \citenamefont {Tiggelman}, \citenamefont {dos Santos~Martins},
  \citenamefont {Dirkse}, \citenamefont {Wehner},\ and\ \citenamefont
  {Hanson}}]{pompili_realization_2021}%
  \BibitemOpen
  \bibfield  {author} {\bibinfo {author} {\bibfnamefont {M.}~\bibnamefont
  {Pompili}}, \bibinfo {author} {\bibfnamefont {S.~L.~N.}\ \bibnamefont
  {Hermans}}, \bibinfo {author} {\bibfnamefont {S.}~\bibnamefont {Baier}},
  \bibinfo {author} {\bibfnamefont {H.~K.~C.}\ \bibnamefont {Beukers}},
  \bibinfo {author} {\bibfnamefont {P.~C.}\ \bibnamefont {Humphreys}}, \bibinfo
  {author} {\bibfnamefont {R.~N.}\ \bibnamefont {Schouten}}, \bibinfo {author}
  {\bibfnamefont {R.~F.~L.}\ \bibnamefont {Vermeulen}}, \bibinfo {author}
  {\bibfnamefont {M.~J.}\ \bibnamefont {Tiggelman}}, \bibinfo {author}
  {\bibfnamefont {L.}~\bibnamefont {dos Santos~Martins}}, \bibinfo {author}
  {\bibfnamefont {B.}~\bibnamefont {Dirkse}}, \bibinfo {author} {\bibfnamefont
  {S.}~\bibnamefont {Wehner}}, \ and\ \bibinfo {author} {\bibfnamefont
  {R.}~\bibnamefont {Hanson}},\ }\bibfield  {title} {\enquote {\bibinfo {title}
  {Realization of a multinode quantum network of remote solid-state qubits},}\
  }\href {\doibase 10.1126/science.abg1919} {\bibfield  {journal} {\bibinfo
  {journal} {Science}\ }\textbf {\bibinfo {volume} {372}},\ \bibinfo {pages}
  {259--264} (\bibinfo {year} {2021})}\BibitemShut {NoStop}%
\bibitem [{\citenamefont {Pompili}\ \emph {et~al.}(2022)\citenamefont
  {Pompili}, \citenamefont {Delle~Donne}, \citenamefont {te~Raa}, \citenamefont
  {van~der Vecht}, \citenamefont {Skrzypczyk}, \citenamefont {Ferreira},
  \citenamefont {de~Kluijver}, \citenamefont {Stolk}, \citenamefont {Hermans},
  \citenamefont {Pawełczak}, \citenamefont {Kozlowski}, \citenamefont
  {Hanson},\ and\ \citenamefont {Wehner}}]{pompili_experimental_2022}%
  \BibitemOpen
  \bibfield  {author} {\bibinfo {author} {\bibfnamefont {M.}~\bibnamefont
  {Pompili}}, \bibinfo {author} {\bibfnamefont {C.}~\bibnamefont
  {Delle~Donne}}, \bibinfo {author} {\bibfnamefont {I.}~\bibnamefont {te~Raa}},
  \bibinfo {author} {\bibfnamefont {B.}~\bibnamefont {van~der Vecht}}, \bibinfo
  {author} {\bibfnamefont {M.}~\bibnamefont {Skrzypczyk}}, \bibinfo {author}
  {\bibfnamefont {G.}~\bibnamefont {Ferreira}}, \bibinfo {author}
  {\bibfnamefont {L.}~\bibnamefont {de~Kluijver}}, \bibinfo {author}
  {\bibfnamefont {A.~J.}\ \bibnamefont {Stolk}}, \bibinfo {author}
  {\bibfnamefont {S.~L.~N.}\ \bibnamefont {Hermans}}, \bibinfo {author}
  {\bibfnamefont {P.}~\bibnamefont {Pawełczak}}, \bibinfo {author}
  {\bibfnamefont {W.}~\bibnamefont {Kozlowski}}, \bibinfo {author}
  {\bibfnamefont {R.}~\bibnamefont {Hanson}}, \ and\ \bibinfo {author}
  {\bibfnamefont {S.}~\bibnamefont {Wehner}},\ }\bibfield  {title} {\enquote
  {\bibinfo {title} {Experimental demonstration of entanglement delivery using
  a quantum network stack},}\ }\href {\doibase 10.1038/s41534-022-00631-2}
  {\bibfield  {journal} {\bibinfo  {journal} {npj Quantum Information}\
  }\textbf {\bibinfo {volume} {8}},\ \bibinfo {pages} {1--10} (\bibinfo {year}
  {2022})}\BibitemShut {NoStop}%
\bibitem [{\citenamefont {Castelletto}\ \emph {et~al.}(2022)\citenamefont
  {Castelletto}, \citenamefont {Peruzzo}, \citenamefont {Bonato}, \citenamefont
  {Johnson}, \citenamefont {Radulaski}, \citenamefont {Ou}, \citenamefont
  {Kaiser},\ and\ \citenamefont {Wrachtrup}}]{castelletto_silicon_2022}%
  \BibitemOpen
  \bibfield  {author} {\bibinfo {author} {\bibfnamefont {S.}~\bibnamefont
  {Castelletto}}, \bibinfo {author} {\bibfnamefont {A.}~\bibnamefont
  {Peruzzo}}, \bibinfo {author} {\bibfnamefont {C.}~\bibnamefont {Bonato}},
  \bibinfo {author} {\bibfnamefont {B.~C.}\ \bibnamefont {Johnson}}, \bibinfo
  {author} {\bibfnamefont {M.}~\bibnamefont {Radulaski}}, \bibinfo {author}
  {\bibfnamefont {H.}~\bibnamefont {Ou}}, \bibinfo {author} {\bibfnamefont
  {F.}~\bibnamefont {Kaiser}}, \ and\ \bibinfo {author} {\bibfnamefont
  {J.}~\bibnamefont {Wrachtrup}},\ }\bibfield  {title} {\enquote {\bibinfo
  {title} {Silicon {Carbide} {Photonics} {Bridging} {Quantum} {Technology}},}\
  }\href {\doibase 10.1021/acsphotonics.1c01775} {\bibfield  {journal}
  {\bibinfo  {journal} {ACS Photonics}\ }\textbf {\bibinfo {volume} {9}},\
  \bibinfo {pages} {1434--1457} (\bibinfo {year} {2022})}\BibitemShut {NoStop}%
\bibitem [{\citenamefont {Sardi}\ \emph {et~al.}(2020)\citenamefont {Sardi},
  \citenamefont {Kornher}, \citenamefont {Widmann}, \citenamefont {Kolesov},
  \citenamefont {Schiller}, \citenamefont {Reindl}, \citenamefont {Hagel},\
  and\ \citenamefont {Wrachtrup}}]{sardi_scalable_2020}%
  \BibitemOpen
  \bibfield  {author} {\bibinfo {author} {\bibfnamefont {F.}~\bibnamefont
  {Sardi}}, \bibinfo {author} {\bibfnamefont {T.}~\bibnamefont {Kornher}},
  \bibinfo {author} {\bibfnamefont {M.}~\bibnamefont {Widmann}}, \bibinfo
  {author} {\bibfnamefont {R.}~\bibnamefont {Kolesov}}, \bibinfo {author}
  {\bibfnamefont {F.}~\bibnamefont {Schiller}}, \bibinfo {author}
  {\bibfnamefont {T.}~\bibnamefont {Reindl}}, \bibinfo {author} {\bibfnamefont
  {M.}~\bibnamefont {Hagel}}, \ and\ \bibinfo {author} {\bibfnamefont
  {J.}~\bibnamefont {Wrachtrup}},\ }\bibfield  {title} {\enquote {\bibinfo
  {title} {Scalable production of solid-immersion lenses for quantum emitters
  in silicon carbide},}\ }\href {\doibase 10.1063/5.0011366} {\bibfield
  {journal} {\bibinfo  {journal} {Applied Physics Letters}\ }\textbf {\bibinfo
  {volume} {117}},\ \bibinfo {pages} {022105} (\bibinfo {year}
  {2020})}\BibitemShut {NoStop}%
\bibitem [{\citenamefont {Widmann}\ \emph {et~al.}(2015)\citenamefont
  {Widmann}, \citenamefont {Lee}, \citenamefont {Rendler}, \citenamefont {Son},
  \citenamefont {Fedder}, \citenamefont {Paik}, \citenamefont {Yang},
  \citenamefont {Zhao}, \citenamefont {Yang}, \citenamefont {Booker},
  \citenamefont {Denisenko}, \citenamefont {Jamali}, \citenamefont
  {Momenzadeh}, \citenamefont {Gerhardt}, \citenamefont {Ohshima},
  \citenamefont {Gali}, \citenamefont {Janzén},\ and\ \citenamefont
  {Wrachtrup}}]{widmann_coherent_2015}%
  \BibitemOpen
  \bibfield  {author} {\bibinfo {author} {\bibfnamefont {M.}~\bibnamefont
  {Widmann}}, \bibinfo {author} {\bibfnamefont {S.-Y.}\ \bibnamefont {Lee}},
  \bibinfo {author} {\bibfnamefont {T.}~\bibnamefont {Rendler}}, \bibinfo
  {author} {\bibfnamefont {N.~T.}\ \bibnamefont {Son}}, \bibinfo {author}
  {\bibfnamefont {H.}~\bibnamefont {Fedder}}, \bibinfo {author} {\bibfnamefont
  {S.}~\bibnamefont {Paik}}, \bibinfo {author} {\bibfnamefont {L.-P.}\
  \bibnamefont {Yang}}, \bibinfo {author} {\bibfnamefont {N.}~\bibnamefont
  {Zhao}}, \bibinfo {author} {\bibfnamefont {S.}~\bibnamefont {Yang}}, \bibinfo
  {author} {\bibfnamefont {I.}~\bibnamefont {Booker}}, \bibinfo {author}
  {\bibfnamefont {A.}~\bibnamefont {Denisenko}}, \bibinfo {author}
  {\bibfnamefont {M.}~\bibnamefont {Jamali}}, \bibinfo {author} {\bibfnamefont
  {S.~A.}\ \bibnamefont {Momenzadeh}}, \bibinfo {author} {\bibfnamefont
  {I.}~\bibnamefont {Gerhardt}}, \bibinfo {author} {\bibfnamefont
  {T.}~\bibnamefont {Ohshima}}, \bibinfo {author} {\bibfnamefont
  {A.}~\bibnamefont {Gali}}, \bibinfo {author} {\bibfnamefont {E.}~\bibnamefont
  {Janzén}}, \ and\ \bibinfo {author} {\bibfnamefont {J.}~\bibnamefont
  {Wrachtrup}},\ }\bibfield  {title} {\enquote {\bibinfo {title} {Coherent
  control of single spins in silicon carbide at room temperature},}\ }\href
  {\doibase 10.1038/nmat4145} {\bibfield  {journal} {\bibinfo  {journal}
  {Nature Materials}\ }\textbf {\bibinfo {volume} {14}},\ \bibinfo {pages}
  {164--168} (\bibinfo {year} {2015})}\BibitemShut {NoStop}%
\bibitem [{\citenamefont {Fu}\ and\ \citenamefont
  {Bryan}(2002)}]{fu_semiconductor_2002}%
  \BibitemOpen
  \bibfield  {author} {\bibinfo {author} {\bibfnamefont {Y.}~\bibnamefont
  {Fu}}\ and\ \bibinfo {author} {\bibfnamefont {N.}~\bibnamefont {Bryan}},\
  }\bibfield  {title} {\enquote {\bibinfo {title} {Semiconductor microlenses
  fabricated by one-step focused ion beam direct writing},}\ }\href {\doibase
  10.1109/66.999597} {\bibfield  {journal} {\bibinfo  {journal} {IEEE
  Transactions on Semiconductor Manufacturing}\ }\textbf {\bibinfo {volume}
  {15}},\ \bibinfo {pages} {229--231} (\bibinfo {year} {2002})}\BibitemShut
  {NoStop}%
\bibitem [{\citenamefont {Nagy}\ \emph {et~al.}(2019)\citenamefont {Nagy},
  \citenamefont {Niethammer}, \citenamefont {Widmann}, \citenamefont {Chen},
  \citenamefont {Udvarhelyi}, \citenamefont {Bonato}, \citenamefont {Hassan},
  \citenamefont {Karhu}, \citenamefont {Ivanov}, \citenamefont {Son},
  \citenamefont {Maze}, \citenamefont {Ohshima}, \citenamefont {Soykal},
  \citenamefont {Gali}, \citenamefont {Lee}, \citenamefont {Kaiser},\ and\
  \citenamefont {Wrachtrup}}]{nagy_high-fidelity_2019}%
  \BibitemOpen
  \bibfield  {author} {\bibinfo {author} {\bibfnamefont {R.}~\bibnamefont
  {Nagy}}, \bibinfo {author} {\bibfnamefont {M.}~\bibnamefont {Niethammer}},
  \bibinfo {author} {\bibfnamefont {M.}~\bibnamefont {Widmann}}, \bibinfo
  {author} {\bibfnamefont {Y.-C.}\ \bibnamefont {Chen}}, \bibinfo {author}
  {\bibfnamefont {P.}~\bibnamefont {Udvarhelyi}}, \bibinfo {author}
  {\bibfnamefont {C.}~\bibnamefont {Bonato}}, \bibinfo {author} {\bibfnamefont
  {J.~U.}\ \bibnamefont {Hassan}}, \bibinfo {author} {\bibfnamefont
  {R.}~\bibnamefont {Karhu}}, \bibinfo {author} {\bibfnamefont {I.~G.}\
  \bibnamefont {Ivanov}}, \bibinfo {author} {\bibfnamefont {N.~T.}\
  \bibnamefont {Son}}, \bibinfo {author} {\bibfnamefont {J.~R.}\ \bibnamefont
  {Maze}}, \bibinfo {author} {\bibfnamefont {T.}~\bibnamefont {Ohshima}},
  \bibinfo {author} {\bibfnamefont {{\"O}.~O.}\ \bibnamefont {Soykal}},
  \bibinfo {author} {\bibfnamefont {{\'A}.}~\bibnamefont {Gali}}, \bibinfo
  {author} {\bibfnamefont {S.-Y.}\ \bibnamefont {Lee}}, \bibinfo {author}
  {\bibfnamefont {F.}~\bibnamefont {Kaiser}}, \ and\ \bibinfo {author}
  {\bibfnamefont {J.}~\bibnamefont {Wrachtrup}},\ }\bibfield  {title} {\enquote
  {\bibinfo {title} {High-fidelity spin and optical control of single
  silicon-vacancy centres in silicon carbide},}\ }\href {\doibase
  10.1038/s41467-019-09873-9} {\bibfield  {journal} {\bibinfo  {journal}
  {Nature Communications}\ }\textbf {\bibinfo {volume} {10}},\ \bibinfo {pages}
  {1954} (\bibinfo {year} {2019})}\BibitemShut {NoStop}%
\bibitem [{\citenamefont {Popovic}, \citenamefont {Sprague},\ and\
  \citenamefont {Connell}(1988)}]{popovic_technique_1988}%
  \BibitemOpen
  \bibfield  {author} {\bibinfo {author} {\bibfnamefont {Z.~D.}\ \bibnamefont
  {Popovic}}, \bibinfo {author} {\bibfnamefont {R.~A.}\ \bibnamefont
  {Sprague}}, \ and\ \bibinfo {author} {\bibfnamefont {G.~A.~N.}\ \bibnamefont
  {Connell}},\ }\bibfield  {title} {\enquote {\bibinfo {title} {Technique for
  monolithic fabrication of microlens arrays},}\ }\href {\doibase
  10.1364/AO.27.001281} {\bibfield  {journal} {\bibinfo  {journal} {Applied
  Optics}\ }\textbf {\bibinfo {volume} {27}},\ \bibinfo {pages} {1281--1284}
  (\bibinfo {year} {1988})}\BibitemShut {NoStop}%
\bibitem [{\citenamefont {Haselbeck}\ \emph {et~al.}(1993)\citenamefont
  {Haselbeck}, \citenamefont {Schreiber}, \citenamefont {Schwider},\ and\
  \citenamefont {Streibl}}]{haselbeck_microlenses_1993}%
  \BibitemOpen
  \bibfield  {author} {\bibinfo {author} {\bibfnamefont {S.}~\bibnamefont
  {Haselbeck}}, \bibinfo {author} {\bibfnamefont {H.}~\bibnamefont
  {Schreiber}}, \bibinfo {author} {\bibfnamefont {J.}~\bibnamefont {Schwider}},
  \ and\ \bibinfo {author} {\bibfnamefont {N.}~\bibnamefont {Streibl}},\
  }\bibfield  {title} {\enquote {\bibinfo {title} {Microlenses fabricated by
  melting a photoresist on a base layer},}\ }\href {\doibase 10.1117/12.135845}
  {\bibfield  {journal} {\bibinfo  {journal} {Optical Engineering}\ }\textbf
  {\bibinfo {volume} {32}},\ \bibinfo {pages} {1322--1324} (\bibinfo {year}
  {1993})}\BibitemShut {NoStop}%
\bibitem [{\citenamefont {Roy}\ \emph {et~al.}(2009)\citenamefont {Roy},
  \citenamefont {Voisin}, \citenamefont {Gravel}, \citenamefont {Peytavi},
  \citenamefont {Boudreau},\ and\ \citenamefont {Veres}}]{roy_microlens_2009}%
  \BibitemOpen
  \bibfield  {author} {\bibinfo {author} {\bibfnamefont {E.}~\bibnamefont
  {Roy}}, \bibinfo {author} {\bibfnamefont {B.}~\bibnamefont {Voisin}},
  \bibinfo {author} {\bibfnamefont {J.-F.}\ \bibnamefont {Gravel}}, \bibinfo
  {author} {\bibfnamefont {R.}~\bibnamefont {Peytavi}}, \bibinfo {author}
  {\bibfnamefont {D.}~\bibnamefont {Boudreau}}, \ and\ \bibinfo {author}
  {\bibfnamefont {T.}~\bibnamefont {Veres}},\ }\bibfield  {title} {\enquote
  {\bibinfo {title} {Microlens array fabrication by enhanced thermal reflow
  process: {Towards} efficient collection of fluorescence light from
  microarrays},}\ }\href {\doibase 10.1016/j.mee.2009.04.001} {\bibfield
  {journal} {\bibinfo  {journal} {Microelectronic Engineering}\ }\textbf
  {\bibinfo {volume} {86}},\ \bibinfo {pages} {2255--2261} (\bibinfo {year}
  {2009})}\BibitemShut {NoStop}%
\bibitem [{\citenamefont {Feng}\ \emph {et~al.}(2012)\citenamefont {Feng},
  \citenamefont {Sihai}, \citenamefont {Huan}, \citenamefont {Yifan},
  \citenamefont {Jianjun},\ and\ \citenamefont
  {Yiqing}}]{feng_fabrication_2012}%
  \BibitemOpen
  \bibfield  {author} {\bibinfo {author} {\bibfnamefont {L.}~\bibnamefont
  {Feng}}, \bibinfo {author} {\bibfnamefont {C.}~\bibnamefont {Sihai}},
  \bibinfo {author} {\bibfnamefont {L.}~\bibnamefont {Huan}}, \bibinfo {author}
  {\bibfnamefont {Z.}~\bibnamefont {Yifan}}, \bibinfo {author} {\bibfnamefont
  {L.}~\bibnamefont {Jianjun}}, \ and\ \bibinfo {author} {\bibfnamefont
  {G.}~\bibnamefont {Yiqing}},\ }\bibfield  {title} {\enquote {\bibinfo {title}
  {Fabrication and characterization of polydimethylsiloxane concave microlens
  array},}\ }\href {\doibase 10.1016/j.optlastec.2011.10.010} {\bibfield
  {journal} {\bibinfo  {journal} {Optics \& Laser Technology}\ }\textbf
  {\bibinfo {volume} {44}},\ \bibinfo {pages} {1054--1059} (\bibinfo {year}
  {2012})}\BibitemShut {NoStop}%
\bibitem [{\citenamefont {Bae}\ \emph {et~al.}(2020)\citenamefont {Bae},
  \citenamefont {Kim}, \citenamefont {Yang}, \citenamefont {Jang},\ and\
  \citenamefont {Jeong}}]{bae_multifocal_2020}%
  \BibitemOpen
  \bibfield  {author} {\bibinfo {author} {\bibfnamefont {S.-I.}\ \bibnamefont
  {Bae}}, \bibinfo {author} {\bibfnamefont {K.}~\bibnamefont {Kim}}, \bibinfo
  {author} {\bibfnamefont {S.}~\bibnamefont {Yang}}, \bibinfo {author}
  {\bibfnamefont {K.-w.}\ \bibnamefont {Jang}}, \ and\ \bibinfo {author}
  {\bibfnamefont {K.-H.}\ \bibnamefont {Jeong}},\ }\bibfield  {title} {\enquote
  {\bibinfo {title} {Multifocal microlens arrays using multilayer
  photolithography},}\ }\href {\doibase 10.1364/OE.388921} {\bibfield
  {journal} {\bibinfo  {journal} {Optics Express}\ }\textbf {\bibinfo {volume}
  {28}},\ \bibinfo {pages} {9082} (\bibinfo {year} {2020})}\BibitemShut
  {NoStop}%
\bibitem [{\citenamefont {Zhou}\ \emph {et~al.}(2016)\citenamefont {Zhou},
  \citenamefont {Peng}, \citenamefont {Peng}, \citenamefont {Zeng},
  \citenamefont {Zhang},\ and\ \citenamefont {Guo}}]{zhou_fabrication_2016}%
  \BibitemOpen
  \bibfield  {author} {\bibinfo {author} {\bibfnamefont {X.}~\bibnamefont
  {Zhou}}, \bibinfo {author} {\bibfnamefont {Y.}~\bibnamefont {Peng}}, \bibinfo
  {author} {\bibfnamefont {R.}~\bibnamefont {Peng}}, \bibinfo {author}
  {\bibfnamefont {X.}~\bibnamefont {Zeng}}, \bibinfo {author} {\bibfnamefont
  {Y.-a.}\ \bibnamefont {Zhang}}, \ and\ \bibinfo {author} {\bibfnamefont
  {T.}~\bibnamefont {Guo}},\ }\bibfield  {title} {\enquote {\bibinfo {title}
  {Fabrication of {Large}-{Scale} {Microlens} {Arrays} {Based} on {Screen}
  {Printing} for {Integral} {Imaging} {3D} {Display}},}\ }\href {\doibase
  10.1021/acsami.6b08278} {\bibfield  {journal} {\bibinfo  {journal} {ACS
  Applied Materials \& Interfaces}\ }\textbf {\bibinfo {volume} {8}},\ \bibinfo
  {pages} {24248--24255} (\bibinfo {year} {2016})}\BibitemShut {NoStop}%
\bibitem [{\citenamefont {Chen}\ \emph {et~al.}(2008)\citenamefont {Chen},
  \citenamefont {Yi}, \citenamefont {Yao}, \citenamefont {Klocke},\ and\
  \citenamefont {Pongs}}]{chen_reflow_2008}%
  \BibitemOpen
  \bibfield  {author} {\bibinfo {author} {\bibfnamefont {Y.}~\bibnamefont
  {Chen}}, \bibinfo {author} {\bibfnamefont {A.~Y.}\ \bibnamefont {Yi}},
  \bibinfo {author} {\bibfnamefont {D.}~\bibnamefont {Yao}}, \bibinfo {author}
  {\bibfnamefont {F.}~\bibnamefont {Klocke}}, \ and\ \bibinfo {author}
  {\bibfnamefont {G.}~\bibnamefont {Pongs}},\ }\bibfield  {title} {\enquote
  {\bibinfo {title} {A reflow process for glass microlens array fabrication by
  use of precision compression molding},}\ }\href {\doibase
  10.1088/0960-1317/18/5/055022} {\bibfield  {journal} {\bibinfo  {journal}
  {Journal of Micromechanics and Microengineering}\ }\textbf {\bibinfo {volume}
  {18}},\ \bibinfo {pages} {055022} (\bibinfo {year} {2008})}\BibitemShut
  {NoStop}%
\bibitem [{\citenamefont {He}\ \emph {et~al.}(2003)\citenamefont {He},
  \citenamefont {Yuan}, \citenamefont {Ngo}, \citenamefont {Bu},\ and\
  \citenamefont {Kudryashov}}]{he_simple_2003}%
  \BibitemOpen
  \bibfield  {author} {\bibinfo {author} {\bibfnamefont {M.}~\bibnamefont
  {He}}, \bibinfo {author} {\bibfnamefont {X.-C.}\ \bibnamefont {Yuan}},
  \bibinfo {author} {\bibfnamefont {N.~Q.}\ \bibnamefont {Ngo}}, \bibinfo
  {author} {\bibfnamefont {J.}~\bibnamefont {Bu}}, \ and\ \bibinfo {author}
  {\bibfnamefont {V.}~\bibnamefont {Kudryashov}},\ }\bibfield  {title}
  {\enquote {\bibinfo {title} {Simple reflow technique for fabrication of a
  microlens array in solgel glass},}\ }\href {\doibase 10.1364/OL.28.000731}
  {\bibfield  {journal} {\bibinfo  {journal} {Optics Letters}\ }\textbf
  {\bibinfo {volume} {28}},\ \bibinfo {pages} {731--733} (\bibinfo {year}
  {2003})}\BibitemShut {NoStop}%
\bibitem [{\citenamefont {Choi}\ \emph {et~al.}(2004)\citenamefont {Choi},
  \citenamefont {Liu}, \citenamefont {Gu}, \citenamefont {McConnell},
  \citenamefont {Girkin}, \citenamefont {Watson},\ and\ \citenamefont
  {Dawson}}]{choi_gan_2004}%
  \BibitemOpen
  \bibfield  {author} {\bibinfo {author} {\bibfnamefont {H.~W.}\ \bibnamefont
  {Choi}}, \bibinfo {author} {\bibfnamefont {C.}~\bibnamefont {Liu}}, \bibinfo
  {author} {\bibfnamefont {E.}~\bibnamefont {Gu}}, \bibinfo {author}
  {\bibfnamefont {G.}~\bibnamefont {McConnell}}, \bibinfo {author}
  {\bibfnamefont {J.~M.}\ \bibnamefont {Girkin}}, \bibinfo {author}
  {\bibfnamefont {I.~M.}\ \bibnamefont {Watson}}, \ and\ \bibinfo {author}
  {\bibfnamefont {M.~D.}\ \bibnamefont {Dawson}},\ }\bibfield  {title}
  {\enquote {\bibinfo {title} {{GaN} micro-light-emitting diode arrays with
  monolithically integrated sapphire microlenses},}\ }\href {\doibase
  10.1063/1.1690876} {\bibfield  {journal} {\bibinfo  {journal} {Applied
  Physics Letters}\ }\textbf {\bibinfo {volume} {84}},\ \bibinfo {pages}
  {2253--2255} (\bibinfo {year} {2004})}\BibitemShut {NoStop}%
\bibitem [{\citenamefont {Park}\ \emph {et~al.}(2001)\citenamefont {Park},
  \citenamefont {Jeon}, \citenamefont {Sung},\ and\ \citenamefont
  {Yeom}}]{park_refractive_2001}%
  \BibitemOpen
  \bibfield  {author} {\bibinfo {author} {\bibfnamefont {S.-H.}\ \bibnamefont
  {Park}}, \bibinfo {author} {\bibfnamefont {H.}~\bibnamefont {Jeon}}, \bibinfo
  {author} {\bibfnamefont {Y.-J.}\ \bibnamefont {Sung}}, \ and\ \bibinfo
  {author} {\bibfnamefont {G.-Y.}\ \bibnamefont {Yeom}},\ }\bibfield  {title}
  {\enquote {\bibinfo {title} {Refractive sapphire microlenses fabricated by
  chlorine-based inductively coupled plasma etching},}\ }\href {\doibase
  10.1364/AO.40.003698} {\bibfield  {journal} {\bibinfo  {journal} {Applied
  Optics}\ }\textbf {\bibinfo {volume} {40}},\ \bibinfo {pages} {3698}
  (\bibinfo {year} {2001})}\BibitemShut {NoStop}%
\bibitem [{\citenamefont {Liu}\ \emph {et~al.}(2016{\natexlab{a}})\citenamefont
  {Liu}, \citenamefont {Herrnsdorf}, \citenamefont {Gu},\ and\ \citenamefont
  {Dawson}}]{liu_control_2016}%
  \BibitemOpen
  \bibfield  {author} {\bibinfo {author} {\bibfnamefont {H.}~\bibnamefont
  {Liu}}, \bibinfo {author} {\bibfnamefont {J.}~\bibnamefont {Herrnsdorf}},
  \bibinfo {author} {\bibfnamefont {E.}~\bibnamefont {Gu}}, \ and\ \bibinfo
  {author} {\bibfnamefont {M.~D.}\ \bibnamefont {Dawson}},\ }\bibfield  {title}
  {\enquote {\bibinfo {title} {Control of edge bulge evolution during
  photoresist reflow and its application to diamond microlens fabrication},}\
  }\href {\doibase 10.1116/1.4943558} {\bibfield  {journal} {\bibinfo
  {journal} {Journal of Vacuum Science \& Technology B, Nanotechnology and
  Microelectronics: Materials, Processing, Measurement, and Phenomena}\
  }\textbf {\bibinfo {volume} {34}},\ \bibinfo {pages} {021602} (\bibinfo
  {year} {2016}{\natexlab{a}})}\BibitemShut {NoStop}%
\bibitem [{\citenamefont {Liu}\ \emph {et~al.}(2016{\natexlab{b}})\citenamefont
  {Liu}, \citenamefont {Reilly}, \citenamefont {Herrnsdorf}, \citenamefont
  {Xie}, \citenamefont {Savitski}, \citenamefont {Kemp}, \citenamefont {Gu},\
  and\ \citenamefont {Dawson}}]{liu_large_2016}%
  \BibitemOpen
  \bibfield  {author} {\bibinfo {author} {\bibfnamefont {H.}~\bibnamefont
  {Liu}}, \bibinfo {author} {\bibfnamefont {S.}~\bibnamefont {Reilly}},
  \bibinfo {author} {\bibfnamefont {J.}~\bibnamefont {Herrnsdorf}}, \bibinfo
  {author} {\bibfnamefont {E.}~\bibnamefont {Xie}}, \bibinfo {author}
  {\bibfnamefont {V.~G.}\ \bibnamefont {Savitski}}, \bibinfo {author}
  {\bibfnamefont {A.~J.}\ \bibnamefont {Kemp}}, \bibinfo {author}
  {\bibfnamefont {E.}~\bibnamefont {Gu}}, \ and\ \bibinfo {author}
  {\bibfnamefont {M.~D.}\ \bibnamefont {Dawson}},\ }\bibfield  {title}
  {\enquote {\bibinfo {title} {Large radius of curvature micro-lenses on single
  crystal diamond for application in monolithic diamond {Raman} lasers},}\
  }\href {\doibase 10.1016/j.diamond.2016.01.016} {\bibfield  {journal}
  {\bibinfo  {journal} {Diamond and Related Materials}\ }\bibinfo {series}
  {Special {Issue} “26th {International} {Conference} on {Diamond} and
  {Carbon} {Materials} – {DCM} 2015”},\ \textbf {\bibinfo {volume} {65}},\
  \bibinfo {pages} {37--41} (\bibinfo {year} {2016}{\natexlab{b}})}\BibitemShut
  {NoStop}%
\bibitem [{\citenamefont {Park}\ \emph {et~al.}(2007)\citenamefont {Park},
  \citenamefont {Kim}, \citenamefont {Hong}, \citenamefont {An},\ and\
  \citenamefont {Oh}}]{park_photoresist_2007}%
  \BibitemOpen
  \bibfield  {author} {\bibinfo {author} {\bibfnamefont {J.-M.}\ \bibnamefont
  {Park}}, \bibinfo {author} {\bibfnamefont {E.-J.}\ \bibnamefont {Kim}},
  \bibinfo {author} {\bibfnamefont {J.-Y.}\ \bibnamefont {Hong}}, \bibinfo
  {author} {\bibfnamefont {I.}~\bibnamefont {An}}, \ and\ \bibinfo {author}
  {\bibfnamefont {H.-K.}\ \bibnamefont {Oh}},\ }\bibfield  {title} {\enquote
  {\bibinfo {title} {Photoresist {Adhesion} {Effect} of {Resist} {Reflow}
  {Process}},}\ }\href {\doibase 10.1143/JJAP.46.5738} {\bibfield  {journal}
  {\bibinfo  {journal} {Japanese Journal of Applied Physics}\ }\textbf
  {\bibinfo {volume} {46}},\ \bibinfo {pages} {5738--5741} (\bibinfo {year}
  {2007})}\BibitemShut {NoStop}%
\bibitem [{\citenamefont {Jucius}\ \emph {et~al.}(2017)\citenamefont {Jucius},
  \citenamefont {Grigaliūnas}, \citenamefont {Lazauskas}, \citenamefont
  {Sapeliauskas}, \citenamefont {Abakevičienė}, \citenamefont {Smetona},\
  and\ \citenamefont {Tamulevičius}}]{jucius_effect_2017}%
  \BibitemOpen
  \bibfield  {author} {\bibinfo {author} {\bibfnamefont {D.}~\bibnamefont
  {Jucius}}, \bibinfo {author} {\bibfnamefont {V.}~\bibnamefont
  {Grigaliūnas}}, \bibinfo {author} {\bibfnamefont {A.}~\bibnamefont
  {Lazauskas}}, \bibinfo {author} {\bibfnamefont {E.}~\bibnamefont
  {Sapeliauskas}}, \bibinfo {author} {\bibfnamefont {B.}~\bibnamefont
  {Abakevičienė}}, \bibinfo {author} {\bibfnamefont {S.}~\bibnamefont
  {Smetona}}, \ and\ \bibinfo {author} {\bibfnamefont {S.}~\bibnamefont
  {Tamulevičius}},\ }\bibfield  {title} {\enquote {\bibinfo {title} {Effect of
  fused silica surface wettability on thermal reflow of polymer microlens
  arrays},}\ }\href {\doibase 10.1007/s00542-016-2975-3} {\bibfield  {journal}
  {\bibinfo  {journal} {Microsystem Technologies}\ }\textbf {\bibinfo {volume}
  {23}},\ \bibinfo {pages} {2193--2206} (\bibinfo {year} {2017})}\BibitemShut
  {NoStop}%
\bibitem [{\citenamefont {Kirchner}\ \emph {et~al.}(2016)\citenamefont
  {Kirchner}, \citenamefont {Guzenko}, \citenamefont {Vartiainen},
  \citenamefont {Chidambaram},\ and\ \citenamefont
  {Schift}}]{kirchner_zep520a_2016}%
  \BibitemOpen
  \bibfield  {author} {\bibinfo {author} {\bibfnamefont {R.}~\bibnamefont
  {Kirchner}}, \bibinfo {author} {\bibfnamefont {V.}~\bibnamefont {Guzenko}},
  \bibinfo {author} {\bibfnamefont {I.}~\bibnamefont {Vartiainen}}, \bibinfo
  {author} {\bibfnamefont {N.}~\bibnamefont {Chidambaram}}, \ and\ \bibinfo
  {author} {\bibfnamefont {H.}~\bibnamefont {Schift}},\ }\bibfield  {title}
  {\enquote {\bibinfo {title} {{ZEP520A} — {A} resist for electron-beam
  grayscale lithography and thermal reflow},}\ }\href {\doibase
  10.1016/j.mee.2016.01.017} {\bibfield  {journal} {\bibinfo  {journal}
  {Microelectronic Engineering}\ }\textbf {\bibinfo {volume} {153}},\ \bibinfo
  {pages} {71--76} (\bibinfo {year} {2016})}\BibitemShut {NoStop}%
\bibitem [{\citenamefont {Xu}, \citenamefont {Zhou},\ and\ \citenamefont
  {Cheung}(2021)}]{xu_fabrication_2021}%
  \BibitemOpen
  \bibfield  {author} {\bibinfo {author} {\bibfnamefont {R.}~\bibnamefont
  {Xu}}, \bibinfo {author} {\bibfnamefont {T.}~\bibnamefont {Zhou}}, \ and\
  \bibinfo {author} {\bibfnamefont {R.}~\bibnamefont {Cheung}},\ }\bibfield
  {title} {\enquote {\bibinfo {title} {Fabrication of {SiC} concave microlens
  array mold based on microspheres self-assembly},}\ }\href {\doibase
  10.1016/j.mee.2020.111481} {\bibfield  {journal} {\bibinfo  {journal}
  {Microelectronic Engineering}\ }\textbf {\bibinfo {volume} {236}},\ \bibinfo
  {pages} {111481} (\bibinfo {year} {2021})}\BibitemShut {NoStop}%
\bibitem [{\citenamefont {Grigaliūnas}\ \emph {et~al.}(2016)\citenamefont
  {Grigaliūnas}, \citenamefont {Lazauskas}, \citenamefont {Jucius},
  \citenamefont {Viržonis}, \citenamefont {Abakevičienė}, \citenamefont
  {Smetona},\ and\ \citenamefont {Tamulevičius}}]{grigaliunas_microlens_2016}%
  \BibitemOpen
  \bibfield  {author} {\bibinfo {author} {\bibfnamefont {V.}~\bibnamefont
  {Grigaliūnas}}, \bibinfo {author} {\bibfnamefont {A.}~\bibnamefont
  {Lazauskas}}, \bibinfo {author} {\bibfnamefont {D.}~\bibnamefont {Jucius}},
  \bibinfo {author} {\bibfnamefont {D.}~\bibnamefont {Viržonis}}, \bibinfo
  {author} {\bibfnamefont {B.}~\bibnamefont {Abakevičienė}}, \bibinfo
  {author} {\bibfnamefont {S.}~\bibnamefont {Smetona}}, \ and\ \bibinfo
  {author} {\bibfnamefont {S.}~\bibnamefont {Tamulevičius}},\ }\bibfield
  {title} {\enquote {\bibinfo {title} {Microlens fabrication by {3D} electron
  beam lithography combined with thermal reflow technique},}\ }\href {\doibase
  10.1016/j.mee.2016.07.003} {\bibfield  {journal} {\bibinfo  {journal}
  {Microelectronic Engineering}\ }\textbf {\bibinfo {volume} {164}},\ \bibinfo
  {pages} {23--29} (\bibinfo {year} {2016})}\BibitemShut {NoStop}%
\bibitem [{\citenamefont {Erjawetz}\ \emph {et~al.}(2022)\citenamefont
  {Erjawetz}, \citenamefont {Collé}, \citenamefont {Ekindorf}, \citenamefont
  {Heyl}, \citenamefont {Ritter}, \citenamefont {Reddy},\ and\ \citenamefont
  {Schift}}]{erjawetz_bend_2022}%
  \BibitemOpen
  \bibfield  {author} {\bibinfo {author} {\bibfnamefont {J.}~\bibnamefont
  {Erjawetz}}, \bibinfo {author} {\bibfnamefont {D.}~\bibnamefont {Collé}},
  \bibinfo {author} {\bibfnamefont {G.}~\bibnamefont {Ekindorf}}, \bibinfo
  {author} {\bibfnamefont {P.}~\bibnamefont {Heyl}}, \bibinfo {author}
  {\bibfnamefont {D.}~\bibnamefont {Ritter}}, \bibinfo {author} {\bibfnamefont
  {A.}~\bibnamefont {Reddy}}, \ and\ \bibinfo {author} {\bibfnamefont
  {H.}~\bibnamefont {Schift}},\ }\bibfield  {title} {\enquote {\bibinfo {title}
  {Bend the curve – {Shape} optimization in laser grayscale direct write
  lithography using a single figure of merit},}\ }\href {\doibase
  10.1016/j.mne.2022.100137} {\bibfield  {journal} {\bibinfo  {journal} {Micro
  and Nano Engineering}\ }\textbf {\bibinfo {volume} {15}},\ \bibinfo {pages}
  {100137} (\bibinfo {year} {2022})}\BibitemShut {NoStop}%
\bibitem [{\citenamefont {Jansen}\ \emph {et~al.}(1996)\citenamefont {Jansen},
  \citenamefont {Gardeniers}, \citenamefont {Boer}, \citenamefont
  {Elwenspoek},\ and\ \citenamefont {Fluitman}}]{jansen_survey_1996}%
  \BibitemOpen
  \bibfield  {author} {\bibinfo {author} {\bibfnamefont {H.}~\bibnamefont
  {Jansen}}, \bibinfo {author} {\bibfnamefont {H.}~\bibnamefont {Gardeniers}},
  \bibinfo {author} {\bibfnamefont {M.~d.}\ \bibnamefont {Boer}}, \bibinfo
  {author} {\bibfnamefont {M.}~\bibnamefont {Elwenspoek}}, \ and\ \bibinfo
  {author} {\bibfnamefont {J.}~\bibnamefont {Fluitman}},\ }\bibfield  {title}
  {\enquote {\bibinfo {title} {A survey on the reactive ion etching of silicon
  in microtechnology},}\ }\href {\doibase 10.1088/0960-1317/6/1/002} {\bibfield
   {journal} {\bibinfo  {journal} {Journal of Micromechanics and
  Microengineering}\ }\textbf {\bibinfo {volume} {6}},\ \bibinfo {pages}
  {14--28} (\bibinfo {year} {1996})}\BibitemShut {NoStop}%
\bibitem [{\citenamefont {Ahn}\ \emph {et~al.}(2004)\citenamefont {Ahn},
  \citenamefont {Han}, \citenamefont {Lee}, \citenamefont {Moon},\ and\
  \citenamefont {Lee}}]{ahn_study_2004}%
  \BibitemOpen
  \bibfield  {author} {\bibinfo {author} {\bibfnamefont {S.~C.}\ \bibnamefont
  {Ahn}}, \bibinfo {author} {\bibfnamefont {S.~Y.}\ \bibnamefont {Han}},
  \bibinfo {author} {\bibfnamefont {J.~L.}\ \bibnamefont {Lee}}, \bibinfo
  {author} {\bibfnamefont {J.~H.}\ \bibnamefont {Moon}}, \ and\ \bibinfo
  {author} {\bibfnamefont {B.~T.}\ \bibnamefont {Lee}},\ }\bibfield  {title}
  {\enquote {\bibinfo {title} {A study on the reactive ion etching of {SiC}
  single crystals using inductively coupled plasma of {SF6}-based gas
  mixtures},}\ }\href {\doibase 10.1007/BF03027370} {\bibfield  {journal}
  {\bibinfo  {journal} {Metals and Materials International}\ }\textbf {\bibinfo
  {volume} {10}},\ \bibinfo {pages} {103--106} (\bibinfo {year}
  {2004})}\BibitemShut {NoStop}%
\bibitem [{\citenamefont {Jiang}\ \emph {et~al.}(2003)\citenamefont {Jiang},
  \citenamefont {Cheung}, \citenamefont {Brown},\ and\ \citenamefont
  {Mount}}]{jiang_inductively_2003}%
  \BibitemOpen
  \bibfield  {author} {\bibinfo {author} {\bibfnamefont {L.}~\bibnamefont
  {Jiang}}, \bibinfo {author} {\bibfnamefont {R.}~\bibnamefont {Cheung}},
  \bibinfo {author} {\bibfnamefont {R.}~\bibnamefont {Brown}}, \ and\ \bibinfo
  {author} {\bibfnamefont {A.}~\bibnamefont {Mount}},\ }\bibfield  {title}
  {\enquote {\bibinfo {title} {Inductively coupled plasma etching of {SiC} in
  {SF6}/{O2} and etch-induced surface chemical bonding modifications},}\ }\href
  {\doibase 10.1063/1.1534908} {\bibfield  {journal} {\bibinfo  {journal}
  {Journal of Applied Physics}\ }\textbf {\bibinfo {volume} {93}},\ \bibinfo
  {pages} {1376--1383} (\bibinfo {year} {2003})}\BibitemShut {NoStop}%
\bibitem [{\citenamefont {Khan}\ \emph {et~al.}(2001)\citenamefont {Khan},
  \citenamefont {Roof}, \citenamefont {Zhou},\ and\ \citenamefont
  {Adesida}}]{khan_etching_2001}%
  \BibitemOpen
  \bibfield  {author} {\bibinfo {author} {\bibfnamefont {F.~A.}\ \bibnamefont
  {Khan}}, \bibinfo {author} {\bibfnamefont {B.}~\bibnamefont {Roof}}, \bibinfo
  {author} {\bibfnamefont {L.}~\bibnamefont {Zhou}}, \ and\ \bibinfo {author}
  {\bibfnamefont {I.}~\bibnamefont {Adesida}},\ }\bibfield  {title} {\enquote
  {\bibinfo {title} {Etching of silicon carbide for device fabrication and
  through via-hole formation},}\ }\href {\doibase 10.1007/s11664-001-0018-y}
  {\bibfield  {journal} {\bibinfo  {journal} {Journal of Electronic Materials}\
  }\textbf {\bibinfo {volume} {30}},\ \bibinfo {pages} {212--219} (\bibinfo
  {year} {2001})}\BibitemShut {NoStop}%
\bibitem [{\citenamefont {Plank}\ \emph {et~al.}(2003)\citenamefont {Plank},
  \citenamefont {Blauw}, \citenamefont {Drift},\ and\ \citenamefont
  {Cheung}}]{plank_etching_2003}%
  \BibitemOpen
  \bibfield  {author} {\bibinfo {author} {\bibfnamefont {N.~O.~V.}\
  \bibnamefont {Plank}}, \bibinfo {author} {\bibfnamefont {M.~A.}\ \bibnamefont
  {Blauw}}, \bibinfo {author} {\bibfnamefont {E.~W. J. M. v.~d.}\ \bibnamefont
  {Drift}}, \ and\ \bibinfo {author} {\bibfnamefont {R.}~\bibnamefont
  {Cheung}},\ }\bibfield  {title} {\enquote {\bibinfo {title} {The etching of
  silicon carbide in inductively coupled {SF} $_{\textrm{6}}$ /{O}
  $_{\textrm{2}}$ plasma},}\ }\href {\doibase 10.1088/0022-3727/36/5/310}
  {\bibfield  {journal} {\bibinfo  {journal} {Journal of Physics D: Applied
  Physics}\ }\textbf {\bibinfo {volume} {36}},\ \bibinfo {pages} {482--487}
  (\bibinfo {year} {2003})}\BibitemShut {NoStop}%
\bibitem [{\citenamefont {Oda}\ \emph {et~al.}(2015)\citenamefont {Oda},
  \citenamefont {Wood}, \citenamefont {Ogiya}, \citenamefont {Miyoshi},\ and\
  \citenamefont {Tsuji}}]{oda_optimizing_2015}%
  \BibitemOpen
  \bibfield  {author} {\bibinfo {author} {\bibfnamefont {H.}~\bibnamefont
  {Oda}}, \bibinfo {author} {\bibfnamefont {P.}~\bibnamefont {Wood}}, \bibinfo
  {author} {\bibfnamefont {H.}~\bibnamefont {Ogiya}}, \bibinfo {author}
  {\bibfnamefont {S.}~\bibnamefont {Miyoshi}}, \ and\ \bibinfo {author}
  {\bibfnamefont {O.}~\bibnamefont {Tsuji}},\ }\bibfield  {title} {\enquote
  {\bibinfo {title} {Optimizing the {SiC} {Plasma} {Etching} {Process} for
  {Manufacturing} {Power} {Devices}},}\ \ }(\bibinfo {address} {Scottsdale,
  Arizona, USA},\ \bibinfo {year} {2015})\BibitemShut {NoStop}%
\bibitem [{\citenamefont {Seok}(2020)}]{seok_micro-trench_2020}%
  \BibitemOpen
  \bibfield  {author} {\bibinfo {author} {\bibfnamefont {O.}~\bibnamefont
  {Seok}},\ }\bibfield  {title} {\enquote {\bibinfo {title} {Micro-trench free
  {4H}-{SiC} etching with improved {SiC}/{SiO2} selectivity using inductively
  coupled {SF6}/{O2}/{Ar} plasma},}\ }\href@noop {} {\bibfield  {journal}
  {\bibinfo  {journal} {Physica Scripta}\ }\textbf {\bibinfo {volume} {95}},\
  \bibinfo {pages} {045606} (\bibinfo {year} {2020})}\BibitemShut {NoStop}%
\bibitem [{\citenamefont {Leerungnawarat}\ \emph {et~al.}(2001)\citenamefont
  {Leerungnawarat}, \citenamefont {Lee}, \citenamefont {Pearton}, \citenamefont
  {Ren},\ and\ \citenamefont {Chu}}]{leerungnawarat_comparison_2001}%
  \BibitemOpen
  \bibfield  {author} {\bibinfo {author} {\bibfnamefont {P.}~\bibnamefont
  {Leerungnawarat}}, \bibinfo {author} {\bibfnamefont {K.~P.}\ \bibnamefont
  {Lee}}, \bibinfo {author} {\bibfnamefont {S.~J.}\ \bibnamefont {Pearton}},
  \bibinfo {author} {\bibfnamefont {F.}~\bibnamefont {Ren}}, \ and\ \bibinfo
  {author} {\bibfnamefont {S.~N.~G.}\ \bibnamefont {Chu}},\ }\bibfield  {title}
  {\enquote {\bibinfo {title} {Comparison of {F}\$\_2\$ {Plasma} {Chemistries}
  for {Deep} {Etching} of {SiC}},}\ }\href@noop {} {\bibfield  {journal}
  {\bibinfo  {journal} {Journal of Electronic Materials}\ }\textbf {\bibinfo
  {volume} {30}},\ \bibinfo {pages} {202--206} (\bibinfo {year}
  {2001})}\BibitemShut {NoStop}%
\bibitem [{\citenamefont {Pavkovich}(1986)}]{pavkovich_proximity_1986}%
  \BibitemOpen
  \bibfield  {author} {\bibinfo {author} {\bibfnamefont {J.~M.}\ \bibnamefont
  {Pavkovich}},\ }\bibfield  {title} {\enquote {\bibinfo {title} {Proximity
  effect correction calculations by the integral equation approximate solution
  method},}\ }\href {\doibase 10.1116/1.583369} {\bibfield  {journal} {\bibinfo
   {journal} {Journal of Vacuum Science \& Technology B: Microelectronics and
  Nanometer Structures}\ }\textbf {\bibinfo {volume} {4}},\ \bibinfo {pages}
  {159} (\bibinfo {year} {1986})}\BibitemShut {NoStop}%
\bibitem [{\citenamefont {Du}\ \emph {et~al.}(2000)\citenamefont {Du},
  \citenamefont {Gao}, \citenamefont {Guo}, \citenamefont {Du}, \citenamefont
  {Qiu},\ and\ \citenamefont {Cui}}]{du_precompensation_2000}%
  \BibitemOpen
  \bibfield  {author} {\bibinfo {author} {\bibfnamefont {J.}~\bibnamefont
  {Du}}, \bibinfo {author} {\bibfnamefont {F.}~\bibnamefont {Gao}}, \bibinfo
  {author} {\bibfnamefont {Y.}~\bibnamefont {Guo}}, \bibinfo {author}
  {\bibfnamefont {C.}~\bibnamefont {Du}}, \bibinfo {author} {\bibfnamefont
  {C.}~\bibnamefont {Qiu}}, \ and\ \bibinfo {author} {\bibfnamefont
  {Z.}~\bibnamefont {Cui}},\ }\bibfield  {title} {\enquote {\bibinfo {title}
  {Precompensation approach for improving the quality of laser direct writing
  patterns by a modified proximity function},}\ }\href {\doibase
  10.1117/1.602426} {\bibfield  {journal} {\bibinfo  {journal} {Optical
  Engineering - OPT ENG}\ }\textbf {\bibinfo {volume} {39}},\ \bibinfo {pages}
  {771--775} (\bibinfo {year} {2000})}\BibitemShut {NoStop}%
\bibitem [{\citenamefont {Ivády}\ \emph {et~al.}(2017)\citenamefont {Ivády},
  \citenamefont {Davidsson}, \citenamefont {Son}, \citenamefont {Ohshima},
  \citenamefont {Abrikosov},\ and\ \citenamefont
  {Gali}}]{ivady_identification_2017}%
  \BibitemOpen
  \bibfield  {author} {\bibinfo {author} {\bibfnamefont {V.}~\bibnamefont
  {Ivády}}, \bibinfo {author} {\bibfnamefont {J.}~\bibnamefont {Davidsson}},
  \bibinfo {author} {\bibfnamefont {N.~T.}\ \bibnamefont {Son}}, \bibinfo
  {author} {\bibfnamefont {T.}~\bibnamefont {Ohshima}}, \bibinfo {author}
  {\bibfnamefont {I.~A.}\ \bibnamefont {Abrikosov}}, \ and\ \bibinfo {author}
  {\bibfnamefont {A.}~\bibnamefont {Gali}},\ }\bibfield  {title} {\enquote
  {\bibinfo {title} {Identification of {Si}-vacancy related room-temperature
  qubits in 4 {H} silicon carbide},}\ }\href {\doibase
  10.1103/PhysRevB.96.161114} {\bibfield  {journal} {\bibinfo  {journal}
  {Physical Review B}\ }\textbf {\bibinfo {volume} {96}},\ \bibinfo {pages}
  {161114} (\bibinfo {year} {2017})}\BibitemShut {NoStop}%
\bibitem [{\citenamefont {Schell}, \citenamefont {Neumer},\ and\ \citenamefont
  {Benson}(2014)}]{schell_numerical_2014}%
  \BibitemOpen
  \bibfield  {author} {\bibinfo {author} {\bibfnamefont {A.~W.}\ \bibnamefont
  {Schell}}, \bibinfo {author} {\bibfnamefont {T.}~\bibnamefont {Neumer}}, \
  and\ \bibinfo {author} {\bibfnamefont {O.}~\bibnamefont {Benson}},\
  }\bibfield  {title} {\enquote {\bibinfo {title} {Numerical analysis of
  efficient light extraction with an elliptical solid immersion lens},}\ }\href
  {\doibase 10.1364/OL.39.004639} {\bibfield  {journal} {\bibinfo  {journal}
  {Optics Letters}\ }\textbf {\bibinfo {volume} {39}},\ \bibinfo {pages} {4639}
  (\bibinfo {year} {2014})}\BibitemShut {NoStop}%
\bibitem [{\citenamefont {Widmann}\ \emph {et~al.}(2019)\citenamefont
  {Widmann}, \citenamefont {Niethammer}, \citenamefont {Fedyanin},
  \citenamefont {Khramtsov}, \citenamefont {Rendler}, \citenamefont {Booker},
  \citenamefont {Ul~Hassan}, \citenamefont {Morioka}, \citenamefont {Chen},
  \citenamefont {Ivanov}, \citenamefont {Son}, \citenamefont {Ohshima},
  \citenamefont {Bockstedte}, \citenamefont {Gali}, \citenamefont {Bonato},
  \citenamefont {Lee},\ and\ \citenamefont
  {Wrachtrup}}]{widmann_electrical_2019}%
  \BibitemOpen
  \bibfield  {author} {\bibinfo {author} {\bibfnamefont {M.}~\bibnamefont
  {Widmann}}, \bibinfo {author} {\bibfnamefont {M.}~\bibnamefont {Niethammer}},
  \bibinfo {author} {\bibfnamefont {D.~Y.}\ \bibnamefont {Fedyanin}}, \bibinfo
  {author} {\bibfnamefont {I.~A.}\ \bibnamefont {Khramtsov}}, \bibinfo {author}
  {\bibfnamefont {T.}~\bibnamefont {Rendler}}, \bibinfo {author} {\bibfnamefont
  {I.~D.}\ \bibnamefont {Booker}}, \bibinfo {author} {\bibfnamefont
  {J.}~\bibnamefont {Ul~Hassan}}, \bibinfo {author} {\bibfnamefont
  {N.}~\bibnamefont {Morioka}}, \bibinfo {author} {\bibfnamefont {Y.-C.}\
  \bibnamefont {Chen}}, \bibinfo {author} {\bibfnamefont {I.~G.}\ \bibnamefont
  {Ivanov}}, \bibinfo {author} {\bibfnamefont {N.~T.}\ \bibnamefont {Son}},
  \bibinfo {author} {\bibfnamefont {T.}~\bibnamefont {Ohshima}}, \bibinfo
  {author} {\bibfnamefont {M.}~\bibnamefont {Bockstedte}}, \bibinfo {author}
  {\bibfnamefont {A.}~\bibnamefont {Gali}}, \bibinfo {author} {\bibfnamefont
  {C.}~\bibnamefont {Bonato}}, \bibinfo {author} {\bibfnamefont {S.-Y.}\
  \bibnamefont {Lee}}, \ and\ \bibinfo {author} {\bibfnamefont
  {J.}~\bibnamefont {Wrachtrup}},\ }\bibfield  {title} {\enquote {\bibinfo
  {title} {Electrical {Charge} {State} {Manipulation} of {Single} {Silicon}
  {Vacancies} in a {Silicon} {Carbide} {Quantum} {Optoelectronic} {Device}},}\
  }\href {\doibase 10.1021/acs.nanolett.9b02774} {\bibfield  {journal}
  {\bibinfo  {journal} {Nano Letters}\ }\textbf {\bibinfo {volume} {19}},\
  \bibinfo {pages} {7173--7180} (\bibinfo {year} {2019})}\BibitemShut {NoStop}%
\bibitem [{\citenamefont {Anderson}\ \emph {et~al.}(2019)\citenamefont
  {Anderson}, \citenamefont {Bourassa}, \citenamefont {Miao}, \citenamefont
  {Wolfowicz}, \citenamefont {Mintun}, \citenamefont {Crook}, \citenamefont
  {Abe}, \citenamefont {Ul~Hassan}, \citenamefont {Son}, \citenamefont
  {Ohshima},\ and\ \citenamefont {Awschalom}}]{anderson_electrical_2019}%
  \BibitemOpen
  \bibfield  {author} {\bibinfo {author} {\bibfnamefont {C.~P.}\ \bibnamefont
  {Anderson}}, \bibinfo {author} {\bibfnamefont {A.}~\bibnamefont {Bourassa}},
  \bibinfo {author} {\bibfnamefont {K.~C.}\ \bibnamefont {Miao}}, \bibinfo
  {author} {\bibfnamefont {G.}~\bibnamefont {Wolfowicz}}, \bibinfo {author}
  {\bibfnamefont {P.~J.}\ \bibnamefont {Mintun}}, \bibinfo {author}
  {\bibfnamefont {A.~L.}\ \bibnamefont {Crook}}, \bibinfo {author}
  {\bibfnamefont {H.}~\bibnamefont {Abe}}, \bibinfo {author} {\bibfnamefont
  {J.}~\bibnamefont {Ul~Hassan}}, \bibinfo {author} {\bibfnamefont {N.~T.}\
  \bibnamefont {Son}}, \bibinfo {author} {\bibfnamefont {T.}~\bibnamefont
  {Ohshima}}, \ and\ \bibinfo {author} {\bibfnamefont {D.~D.}\ \bibnamefont
  {Awschalom}},\ }\bibfield  {title} {\enquote {\bibinfo {title} {Electrical
  and optical control of single spins integrated in scalable semiconductor
  devices},}\ }\href {\doibase 10.1126/science.aax9406} {\bibfield  {journal}
  {\bibinfo  {journal} {Science}\ }\textbf {\bibinfo {volume} {366}},\ \bibinfo
  {pages} {1225--1230} (\bibinfo {year} {2019})}\BibitemShut {NoStop}%
\bibitem [{\citenamefont {Elshaari}\ \emph {et~al.}(2020)\citenamefont
  {Elshaari}, \citenamefont {Pernice}, \citenamefont {Srinivasan},
  \citenamefont {Benson},\ and\ \citenamefont
  {Zwiller}}]{elshaari_hybrid_2020}%
  \BibitemOpen
  \bibfield  {author} {\bibinfo {author} {\bibfnamefont {A.~W.}\ \bibnamefont
  {Elshaari}}, \bibinfo {author} {\bibfnamefont {W.}~\bibnamefont {Pernice}},
  \bibinfo {author} {\bibfnamefont {K.}~\bibnamefont {Srinivasan}}, \bibinfo
  {author} {\bibfnamefont {O.}~\bibnamefont {Benson}}, \ and\ \bibinfo {author}
  {\bibfnamefont {V.}~\bibnamefont {Zwiller}},\ }\bibfield  {title} {\enquote
  {\bibinfo {title} {Hybrid integrated quantum photonic circuits},}\ }\href
  {\doibase 10.1038/s41566-020-0609-x} {\bibfield  {journal} {\bibinfo
  {journal} {Nature photonics}\ }\textbf {\bibinfo {volume} {14}},\ \bibinfo
  {pages} {10.1038/s41566--020--0609--x} (\bibinfo {year} {2020})}\BibitemShut
  {NoStop}%
\bibitem [{\citenamefont {Yoo}(2021)}]{yoo_hybrid_2021}%
  \BibitemOpen
  \bibfield  {author} {\bibinfo {author} {\bibfnamefont {S.~J.~B.}\
  \bibnamefont {Yoo}},\ }\bibfield  {title} {\enquote {\bibinfo {title} {Hybrid
  integrated photonic platforms: opinion},}\ }\href {\doibase
  10.1364/OME.438778} {\bibfield  {journal} {\bibinfo  {journal} {Optical
  Materials Express}\ }\textbf {\bibinfo {volume} {11}},\ \bibinfo {pages}
  {3528--3534} (\bibinfo {year} {2021})}\BibitemShut {NoStop}%
\bibitem [{\citenamefont {Toyoda}, \citenamefont {Komiya},\ and\ \citenamefont
  {Itakura}(1980)}]{toyoda_etching_1980}%
  \BibitemOpen
  \bibfield  {author} {\bibinfo {author} {\bibfnamefont {H.}~\bibnamefont
  {Toyoda}}, \bibinfo {author} {\bibfnamefont {H.}~\bibnamefont {Komiya}}, \
  and\ \bibinfo {author} {\bibfnamefont {H.}~\bibnamefont {Itakura}},\
  }\bibfield  {title} {\enquote {\bibinfo {title} {Etching characteristics of
  {SiO2} in {CHF3} gas plasma},}\ }\href {\doibase 10.1007/BF02652937}
  {\bibfield  {journal} {\bibinfo  {journal} {Journal of Electronic Materials}\
  }\textbf {\bibinfo {volume} {9}},\ \bibinfo {pages} {569--584} (\bibinfo
  {year} {1980})}\BibitemShut {NoStop}%
\bibitem [{\citenamefont {Oehrlein}\ \emph {et~al.}(1994)\citenamefont
  {Oehrlein}, \citenamefont {Zhang}, \citenamefont {Vender},\ and\
  \citenamefont {Joubert}}]{oehrlein_fluorocarbon_1994}%
  \BibitemOpen
  \bibfield  {author} {\bibinfo {author} {\bibfnamefont {G.~S.}\ \bibnamefont
  {Oehrlein}}, \bibinfo {author} {\bibfnamefont {Y.}~\bibnamefont {Zhang}},
  \bibinfo {author} {\bibfnamefont {D.}~\bibnamefont {Vender}}, \ and\ \bibinfo
  {author} {\bibfnamefont {O.}~\bibnamefont {Joubert}},\ }\bibfield  {title}
  {\enquote {\bibinfo {title} {Fluorocarbon high‐density plasmas. {II}.
  {Silicon} dioxide and silicon etching using {CF4} and {CHF3}},}\ }\href
  {\doibase 10.1116/1.578877} {\bibfield  {journal} {\bibinfo  {journal}
  {Journal of Vacuum Science \& Technology A}\ }\textbf {\bibinfo {volume}
  {12}},\ \bibinfo {pages} {333--344} (\bibinfo {year} {1994})}\BibitemShut
  {NoStop}%
\bibitem [{\citenamefont {Wang}\ \emph {et~al.}(1998)\citenamefont {Wang},
  \citenamefont {Lambers}, \citenamefont {Pearton}, \citenamefont {Ostling},
  \citenamefont {Zetterling}, \citenamefont {Grow}, \citenamefont {Ren},\ and\
  \citenamefont {Shul}}]{wang_ICP_1998}%
  \BibitemOpen
  \bibfield  {author} {\bibinfo {author} {\bibfnamefont {J.~J.}\ \bibnamefont
  {Wang}}, \bibinfo {author} {\bibfnamefont {E.~S.}\ \bibnamefont {Lambers}},
  \bibinfo {author} {\bibfnamefont {S.~J.}\ \bibnamefont {Pearton}}, \bibinfo
  {author} {\bibfnamefont {M.}~\bibnamefont {Ostling}}, \bibinfo {author}
  {\bibfnamefont {C.~M.}\ \bibnamefont {Zetterling}}, \bibinfo {author}
  {\bibfnamefont {J.~M.}\ \bibnamefont {Grow}}, \bibinfo {author}
  {\bibfnamefont {F.}~\bibnamefont {Ren}}, \ and\ \bibinfo {author}
  {\bibfnamefont {R.~J.}\ \bibnamefont {Shul}},\ }\bibfield  {title} {\enquote
  {\bibinfo {title} {{ICP} etching of {SiC}},}\ }\href {\doibase
  10.1016/S0038-1101(98)00226-3} {\bibfield  {journal} {\bibinfo  {journal}
  {Solid-State Electronics}\ }\textbf {\bibinfo {volume} {42}},\ \bibinfo
  {pages} {2283--2288} (\bibinfo {year} {1998})}\BibitemShut {NoStop}%
\bibitem [{\citenamefont {Ozgur}, \citenamefont {Pedersen},\ and\ \citenamefont
  {Huff}(2018)}]{ozgur_comparison_2018}%
  \BibitemOpen
  \bibfield  {author} {\bibinfo {author} {\bibfnamefont {M.}~\bibnamefont
  {Ozgur}}, \bibinfo {author} {\bibfnamefont {M.}~\bibnamefont {Pedersen}}, \
  and\ \bibinfo {author} {\bibfnamefont {M.}~\bibnamefont {Huff}},\ }\bibfield
  {title} {\enquote {\bibinfo {title} {Comparison of the {Etch} {Mask}
  {Selectivity} of {Nickel} and {Copper} for a {Deep}, {Anisotropic} {Plasma}
  {Etching} {Process} of {Silicon} {Carbide} ({SiC})},}\ }\href {\doibase
  10.1149/2.0121802jss} {\bibfield  {journal} {\bibinfo  {journal} {ECS Journal
  of Solid State Science and Technology}\ }\textbf {\bibinfo {volume} {7}},\
  \bibinfo {pages} {P55} (\bibinfo {year} {2018})}\BibitemShut {NoStop}%
\end{thebibliography}%

\section{Supplementary Information}

\subsection{Grayscale hard-mask fabrication}
\label{app:Hardmask}

In this section, we will discuss the importance of the grayscale hard-mask, and hence the two-step etching process, to achieving the desired outcome of hemispherical solid immersion lenses. 

\begin{figure} [t]
    \centering
    \includegraphics{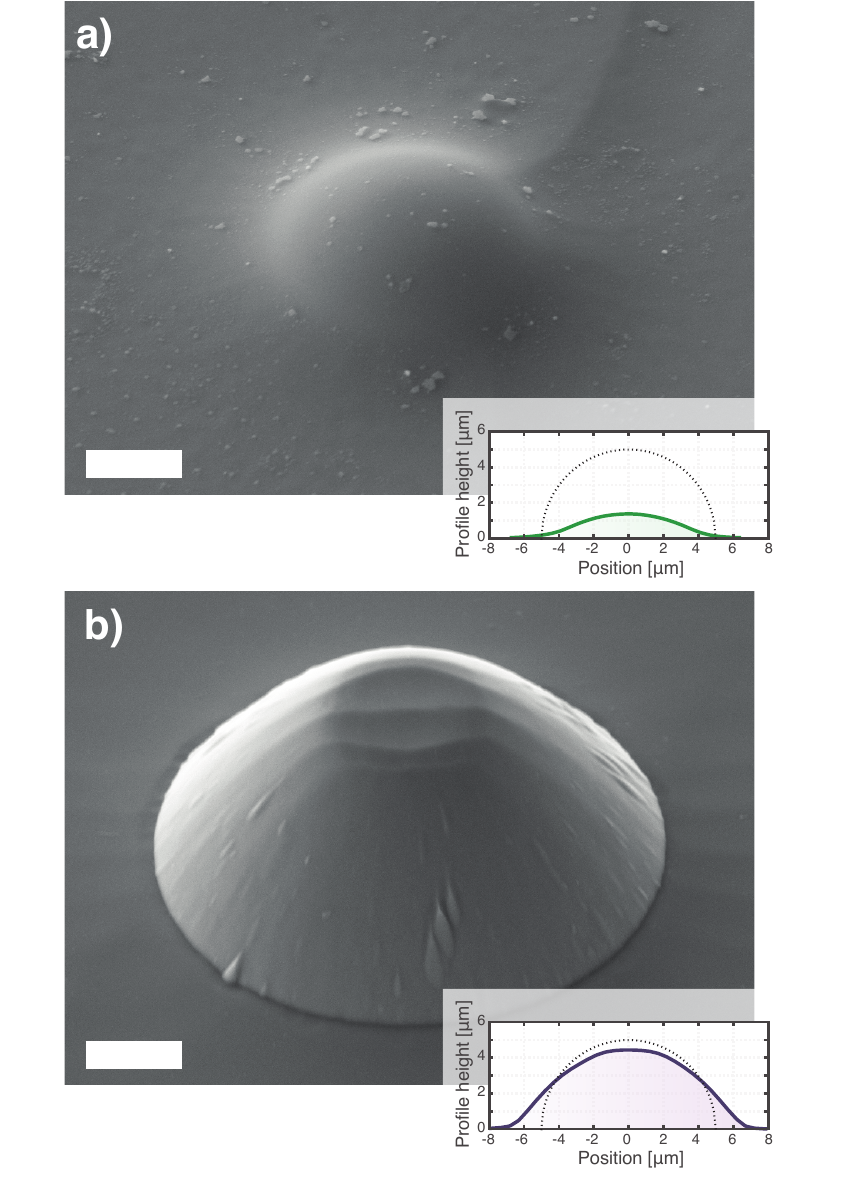}
    \caption{SEMs and associated profiles measured with a stylus profilometer for lenses in silicon carbide fabricated using (a) a single, direct etch using a photoresist mask and (b) a two-step dry etch using an intermediate silica hard mask. Each inset also shows the profile of a 5~$\mu$m radius hemisphere, for reference. Scale bars = 2~$\mu$m.}
    \label{Fig5}
\end{figure}

As mentioned in the main text, conventional grayscale lithography imposes two limitations to the achievable quality and aspect ratios of etched structures. Firstly, the grayscale resist is sensitive to heating during the course of etching, resulting in reflow and bubbling of the resist for etch parameters that are too harsh. Secondly, the optimal etch recipe for SiC requires a proportion of oxygen gas to effectively remove the carbon in the material~\cite{khan_etching_2001, jiang_inductively_2003, ahn_study_2004}, however oxygen also effectively etches organics including the grayscale photoresist mask, leading to low etch selectivity of only around 0.5, even after multiple optimisation attempts (i.e. the final device height is half that of the photoresist mask).  Figure \ref{Fig5}(a) shows a scanning electron micrograph and associated height profile of a typical lens after SiC etch with photoresist mask.

One approach to achieve the desired shape of 5~$\mu$m-radius lenses is to increase the thickness of the resist mask. This allows taller structures in principle, however thicker resists have worse patterning resolution due to the aggravation of proximity effects and defocusing of the laser spot. 

Alternatively, the grayscale hard-mask process introduces a silica (SiO$_2$) layer to decouple etching of the photoresist mask and SiC substrate. This process relies on the comparatively gentle and high-selectivity etch properties of trifluoromethane (CHF$_3$) gas to etch SiO$_2$ with a photoresist mask. In this etch step, it is not only possible to achieve a resist-to-SiO$_2$ etch selectivity of up to $\sim 5-10$~\cite{toyoda_etching_1980, oehrlein_fluorocarbon_1994}, but the etch is sufficiently gentle to minimize reflow and bubbling effects on the photoresist. However, this selectivity is greater than what is needed for our process, where a selectivity close to 1 is ideal to maximize the fidelity of transferred shapes. Therefore additional oxygen can be introduced into the etch chemistry to systematically tune the selectivity to be close to 1, as oxygen etches the photoresist but not the SiO$_2$ hard-mask.

Once this first etch has been accomplished, the same resistance of SiO$_2$ to etching with oxygen is used in the second etch step, where it acts as a mask for etching the SiC substrate. SiO$_2$ is not usually considered as a hard-mask for SiC due to its limited selectivity compared to metal masks such as Ni, Cr or Cu~\cite{wang_ICP_1998, ahn_study_2004, ozgur_comparison_2018}, however at the ideal SiC-etching chemistry a SiO$_2$-to-SiC selectivity of $\sim$5-10 is again achievable~\cite{seok_micro-trench_2020, oda_optimizing_2015}. A similar approach to reducing the selectivity can be applied by introducing CHF$_3$ gas, which etches SiO$_2$ preferentially to SiC, to reach our targeted selectivity of around unity. More importantly, the grayscale hard-mask is robust to temperature rises during the etch process so that the desired lens shapes can be accurately transferred even for harsher etch parameters. The total attained selectivity of the process was therefore raised to a value of 1.2 with the inclusion of the SiO$_2$ grayscale hard-mask, with the ability to tune this higher or lower through the introduction of mask etchants in each step.

The improved aspect ratios possible through the grayscale hard-mask fabrication process is shown in Fig. \ref{Fig5}(b). In the scanning electron micrograph, the slightly flattened top of the lens resulted from underexposure of the resist during development, so that the resist was at full height near the top of the shape (i.e. did not develop at all), however as can be seen in the height profile, the lens is still a reasonable fit for the desired hemispherical shape (dotted line).

\subsection{Experimental Setup and Characterization}
\label{app:Setup}

\begin{figure} [htb]
    \centering
    \includegraphics{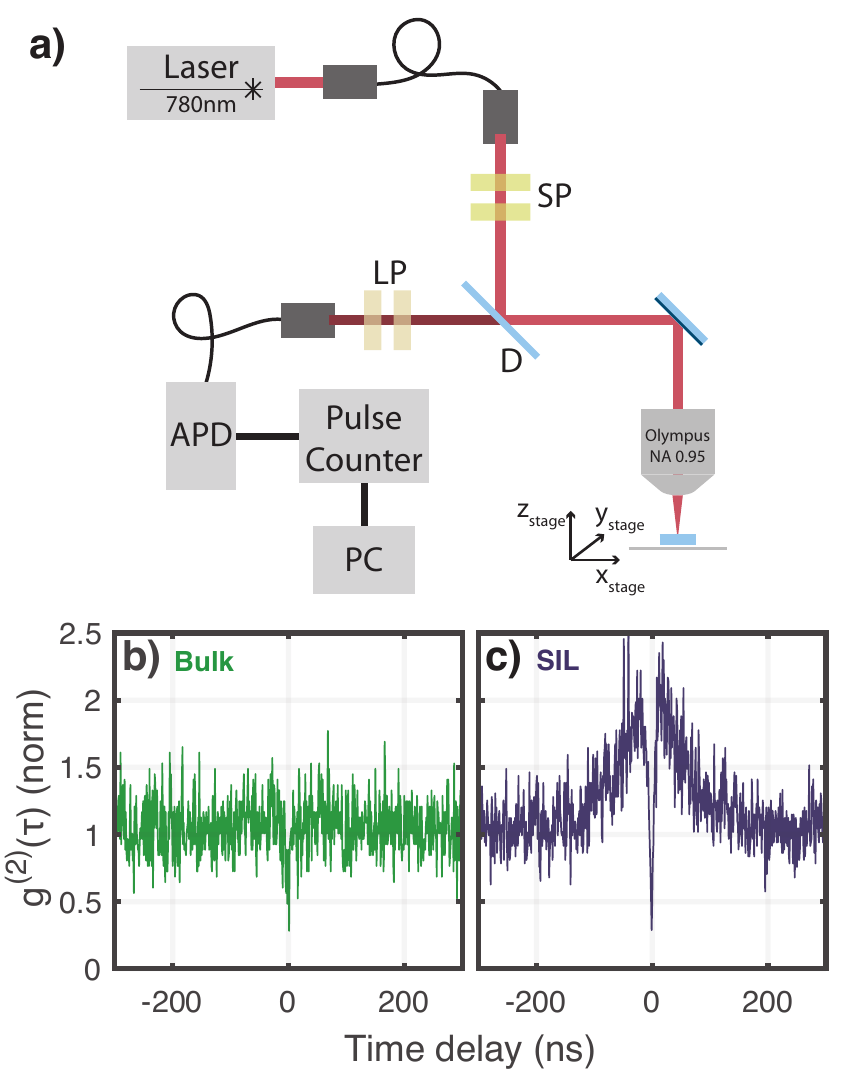}
    \caption{(a) Confocal microscope setup used to characterize fabricated lenses. A 780nm excitation laser is focused onto the sample using a 0.95 NA objective, and resulting photoluminescence from V$_{\text{Si}}$ centers is collected on an avalanche photodiode (APD) single-photon counting module. SP = Short Pass; LP = Long Pass; D = Dichroic. See Section \ref{app:Setup} for a detailed description of the components used. (b) Autocorrelation (g$^{(2)}(\tau)$) plot of a V\textsubscript{Si} center in bulk SiC material, taken at an input power of 1~mW. The dip value below g$^{(2)}(0) = 0.5$ confirms that it is a single photon emitter. (c) The corresponding g$^{(2)}(t)$ plot of a V$_{\text{Si}}$ center taken beneath a lens, displaying the same depth of the autocorrelation dip at zero delay as for the bulk emitter.}
    \label{Fig6}
\end{figure}

We benchmark the lens performance with a custom-built confocal setup at room temperature, measuring the photoluminescence enhancement for single silicon vacancy (V\textsubscript{Si}) centers in SiC. A schematic of the setup is shown in Figure \ref{Fig6}(a). A 780nm laser (Toptica DL Pro) is injected into the excitation arm, filtered using 800nm shortpass filters (Thorlabs FES800) and reflected from a dichroic element (Semrock BrightLine FF801-Di02-25x36) into the common arm of the microscope. From here, a periscope mirror directs the beam into a 0.95 NA air objective (Nikon CF Plan 100X/0.95) to focus on the sample. The collected photoluminescence is collimated by the objective and transmitted through the dichroic element to the collection arm. Here the beam is further filtered using 850nm longpass filters to remove stray laser light (Semrock Versachrome Edge TLP01-887-25x36 and Thorlabs FEL850) and collected into a fibre (SMF28). The fibre is connected to an avalanche photodetector (Excelitas SPCM-900-14-FC) to detect single photons, and a Swabian TimeTagger Ultra is used to count the resultant electrical pulses. Sample scanning is performed by a 3-axis piezoelectric translation stage (Newport NP100SG) mounted on a manual 2-axis stage for coarse positioning. Coarse positioning in Z is provided by a single-axis manual translation stage, on which the objective is mounted.

To ensure that the features observed in photoluminescence maps (e.g. Fig. \ref{Fig4}(a,b)) were indeed single V\textsubscript{Si} emitters, photon correlation measurements were taken both for features within the bulk (Fig. \ref{Fig6}(b)) and under a lens (Fig. \ref{Fig6}(c)). While both plots are taken with an input power of 0.8~mW, the enhancement of the lens leads to both an increase in the excitation intensity experienced by the emitter and an improvement in the collection efficiency of the emitted light, leading to the large bunching 'shoulders' observed in Fig.~\ref{Fig6}(c). The autocorrelation dip at zero delay reaches to the virtually same depth for both emitters, namely $g_{\text{bulk}}^{(2)}(0) = 0.282$ and $g_{\text{SIL}}^{(2)}(0) = 0.287$ for the emitter in the bulk and beneath the lens, respectively, so that they can be readily compared. As the autocorrelation dips observed for both cases also reach below a normalized value of $g^{(2)}(0) = 0.5$,  it is also confirmed that they are single photon sources. 

\end{document}